\documentclass{nature}

\usepackage[T1]{fontenc}
\usepackage{amssymb}
\usepackage{rotating}
\usepackage{gensymb}
   \usepackage{graphicx}
 \usepackage{hyperref}
\RequirePackage{lineno}
\usepackage{txfonts}

\usepackage{float}

\usepackage[usenames]{color}

\usepackage{gensymb}

\newcommand{\be} {\begin{equation}}

\newcommand{\bc}{\begin{center}}
\newcommand{\ec}{\end{center}}
\def\ltsima{$\; \buildrel < \over \sim \;$}
\def\lsim{\lower.5ex\hbox{\ltsima}}
\def\loe{\lower.5ex\hbox{\ltsima}}
\def\gtsima{$\; \buildrel > \over \sim \;$}
\def\gsim{\lower.5ex\hbox{\gtsima}}
\def\goe{\lower.5ex\hbox{\gtsima}}

\def\sss {SS 433}
\def\jj {Fermi J1913+0515}

\def\psrj {PSR\,J1907$+$0602}

\def\ltsima{$\; \buildrel < \over \sim \;$}
\def\lsim{\lower.5ex\hbox{\ltsima}}
\def\loe{\lower.5ex\hbox{\ltsima}}
\def\gtsima{$\; \buildrel > \over \sim \;$}
\def\gsim{\lower.5ex\hbox{\gtsima}}
\def\goe{\lower.5ex\hbox{\gtsima}}

\def\ergscm2 {erg\,s$^{-1}$cm$^{-2}$}

\def\cm2 {cm$^{-2}$}

    {\pagebreak\global\pdfpageattr\expandafter{\the\pdfpageattr/Rotate 90}}%
    {\pagebreak\global\pdfpageattr\expandafter{\the\pdfpageattr/Rotate 0}}%

\title{Gamma-ray heartbeat powered by the microquasar \sss\/}

\author{Jian Li$^{1\dag}$, Diego F. Torres$^{2,3,4}$, Ruo-Yu Liu$^{5,6}$, Matthew Kerr$^{7}$, Emma de O\~na Wilhelmi$^{1,2}$, \& Yang Su$^{8}$}

\makeatletter
\let\saved@includegraphics\includegraphics
\AtBeginDocument{\let\includegraphics\saved@includegraphics}

\makeatother

\begin{document}

\maketitle

\begin{affiliations}
\item Deutsches Elektronen-Synchrotron DESY, D-15738 Zeuthen, Germany\quad $^\dag$jian.li@desy.de
\item Institute of Space Sciences (ICE, CSIC), Campus UAB, Carrer de Magrans s/n, 08193 Barcelona, Spain
\item Instituci\'o Catalana de Recerca i Estudis Avan\c{c}ats (ICREA), E-08010 Barcelona, Spain
\item Institut d'Estudis Espacials de Catalunya (IEEC), 08034 Barcelona, Spain
\item School of Astronomy and Space Science, Nanjing University, Nanjing, China
\item Key Laboratory of Modern Astronomy and Astrophysics (Nanjing University), Ministry of Education, Nanjing, China
\item Space Science Division, Naval Research Laboratory, Washington, DC 20375, USA
\item Purple Mountain Observatory and Key Laboratory of Radio Astronomy, Chinese Academy of Sciences, Nanjing 210034, China

\end{affiliations}

\textbf{
Microquasars, the local siblings of extragalactic quasars, are binary systems comprising a compact object and a companion star.
By accreting matter from their companions, microquasars launch powerful winds and jets, influencing the interstellar environment around them.
Steady gamma-ray emission is expected to rise from their central objects, or from interactions between their outflows and the surrounding medium.
The latter prediction was recently confirmed with the detection of \sss\,\cite{HAWC2018} at high (TeV) energies.
In this report, we analyze more than ten years of GeV gamma-ray data from the \emph{Fermi} Gamma-ray Space Telescope on this source.
{ {Detailed scrutiny of the data reveal emission
in the SS 433 vicinity, co-spatial with a gas enhancement, and hints for emission
possibly associated with a terminal lobe of one of the jets.
Both gamma-ray excesses are relatively far from the central binary, and
the former shows evidence for a periodic variation at the precessional period of \sss, linking
it with the microquasar.}}
This result challenges obvious interpretations and is unexpected from any previously published theoretical models.
It provides us with a chance to unveil the particle transport from \sss\ and to probe the structure of the local magnetic field in its vicinity.
}
%%%%%%%%%%%%%%%%%

\sss\/ is a unique Galactic microquasar containing a compact object, most likely a black hole of  $\sim$10--20 M$_{\odot}$ orbiting a $\sim$30 M$_{\odot}$
A3-7 supergiant star with an orbital period of 13.082~days\cite{Fabrika2004}.
The rate of mass transfer from the companion is determined from the analysis of optical lines\cite{Shklovskii1981} and is thought to be as high as
$10^{-4}$ M$_\odot$ yr$^{-1}$, which is orders of magnitude larger than the Eddington limit.
This steady super-critical accretion state powers highly collimated jets of plasma and mass-loaded non-polar outflows at a similar level\cite{Fabian1979,Margon1979,Middleton2018}, with
kinetic powers exceeding $\gtrsim$10$^{39}$~erg~s$^{-1}$.
The jets appear to inflate the W50 nebula surrounding \sss\cite{Dubner1998}, and perhaps also the H\,{\sc i} shell-like structure seen
on an even larger scale\cite{Su2018}.
Jets and outflows in \sss\/ eject matter at relativistic speeds, $\sim$0.2c\cite{Migliari2002,Middleton2018}, while precessing with a period of 162.250 days\cite{Davydov2008}.
This timing signature is
explained by the periodic pull of the giant secondary star, moving the accretion disk and its outflows in solidarity.
Doppler shifts of H and He lines in the optical as well as of highly ionized Fe lines in the X-rays  indicate relativistic baryon content in the jets\cite{Marshall2002,Migliari2002}; whereas
{knots seen at radio frequency indicate} the existence of relativistic electrons\cite{Vermeulen1987}.

Synchrotron emission in radio and X-ray bands is observed from the jet termination lobes\cite{Dubner1998, Safi-Harb1997}.
Recently, very high energy gamma rays ($>25$ TeV) were also observed at these positions by HAWC, with a likely origin from inverse-Compton scattering between
locally-accelerated electrons and cosmic microwave background radiation\cite{HAWC2018}.
GeV emission would also be expected from the same leptonic processes at the lobes of \sss, although at a level that would be challenging to detect with the \emph{Fermi}-Large Area Telescope (LAT)~\cite{HAWC2018}.
Nonetheless, the existence of baryons in the jets has also promoted models in which gamma-ray emission can rise hadronically at the jet base (see, e.g.,\cite{Reynoso2008}), and/or in interactions between
molecular clouds and cosmic rays that diffuse away from the accelerating region (see, e.g.,\cite{Bosch2005}). We come back to these ideas below, in the context of our findings.

Searches for GeV emission from \sss\/ have thus been a subject of strong interest, and a number of studies using \emph{Fermi}-LAT data have arrived at inconsistent conclusions\cite{Bordas2015, Xing2019, Rasul2019, Sun2019}.
However, as we detail in the \emph{Methods} section, these studies lacked a proper treatment for the contamination produced from nearby sources, in particular from the
pulsar  \psrj, and are thus at risk of systematic biases.

\begin{center}
\begin{figure*}
\centering
\vspace{-2.5cm}
\includegraphics[scale=0.472]{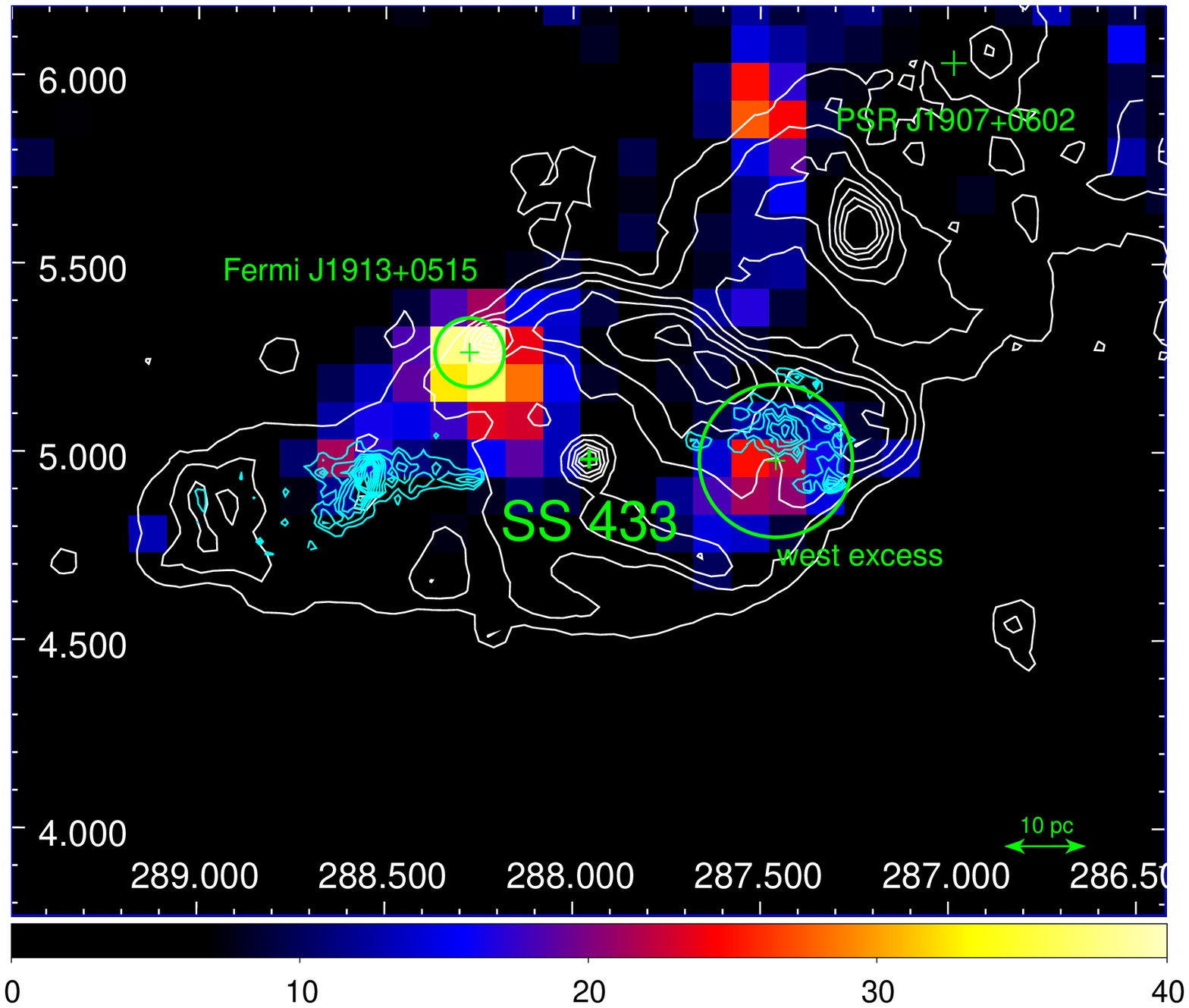} \\
\includegraphics[scale=0.472]{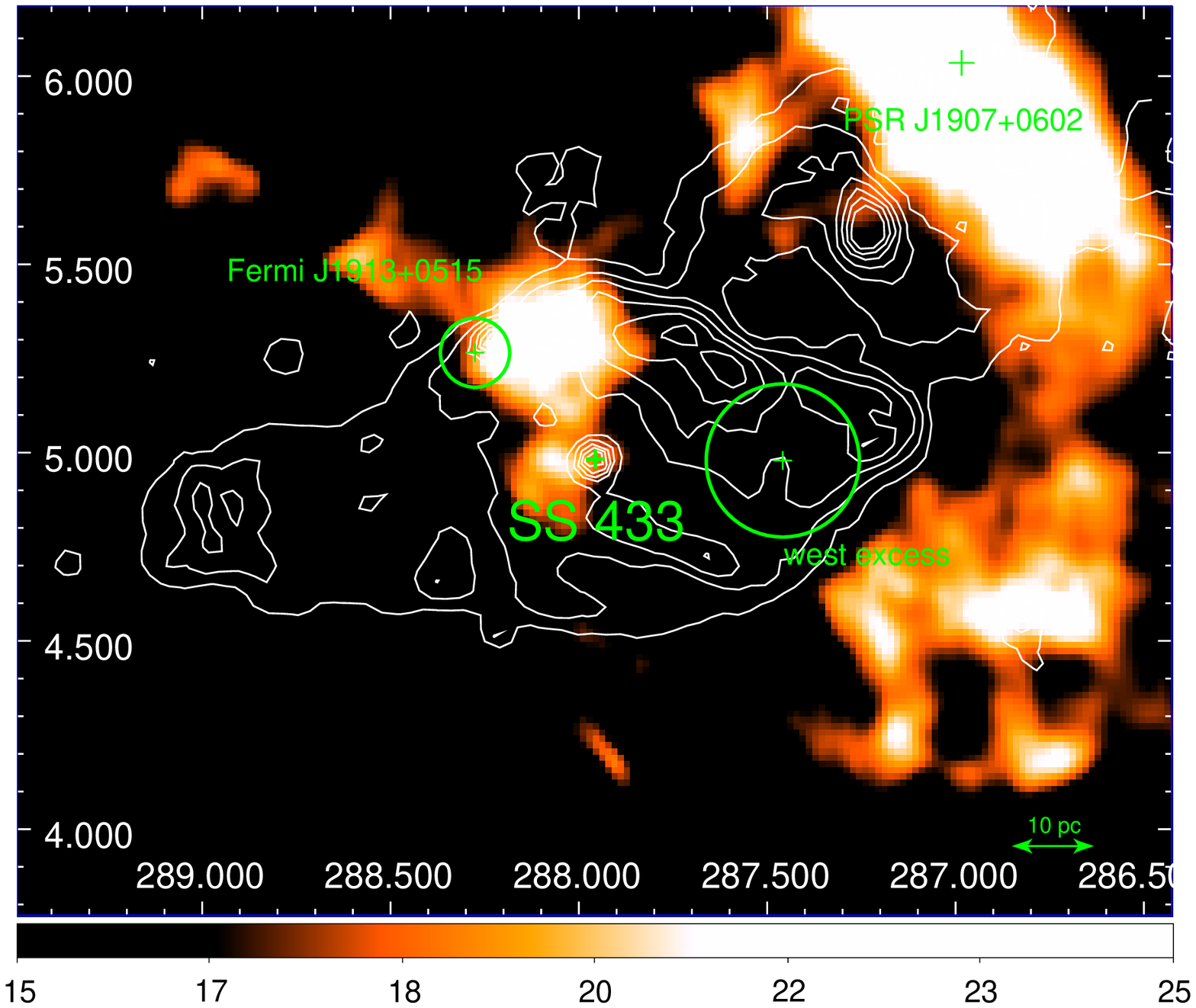}
\caption{Gamma-ray and atomic cloud images of the \sss\/ region.
{\bf Top:} \emph{Fermi}-LAT TS map of \sss\/ region in 0.1--300 GeV during the off-peak phase of \psrj\/.
Background sources have been modelled and subtracted (see \emph{Methods}).
The color scale shows the Test Statistic ($TS$), the square root of which gives an approximate detection significance.
The 95\% confidence level circle of the positions of \jj\ and west excess are shown in green.
The white contours show the radio continuum emission from the Effelsberg 11 cm survey.
Cyan contours show the X-ray emission measured by ROSAT.
 {\bf Bottom:}
Arecibo H\,{\sc i} emission integrated in the interval 65-82 km s$^{-1}$ in units of K km/s.
The image has been scaled by $sin\left| b \right|$ ($b$ is Galactic latitude) to enhance the features far from the Galactic plane\cite{Su2018}.
The x and y axes are R.A. and decl. (J2000, degrees).}

\label{tsmap}
\end{figure*}
\end{center}

Using 10.5 years of \emph{Fermi}-LAT data, we carried out a deep search for gamma-ray emission related to \sss\/ in the 100 MeV -- 300 GeV band during the off-peak phase of \psrj\/.
Full details of our search are described in the \emph{Methods} section.
We detected two GeV excesses near \sss, neither at the position of the central compact object.
These excesses are shown in Figure \ref{tsmap} ({top} panel), together with the radio morphology of the W50 nebula and X-ray contours of the lobes.
There is a GeV excess (hereafter referred to as \jj\/, at { R.A. = 288.28$\degree$$\pm$0.04$\degree$, decl.= 5.27$\degree$$\pm$0.04$\degree$}) {that lies adjacent to} the X-ray contours of the \sss\/ east lobe but does not overlap with them.
\jj\/ is spatially consistent with the \emph{Fermi}-LAT 8-year Point Source List (FL8Y) gamma-ray source FL8Y J1913.3+0515.
No gamma-ray source is found at this location in the \textit{Fermi} Large Area Telescope Fourth Source Catalog (4FGL; see \emph{Methods}).
Assuming a power-law spectral shape ($dN/dE=N_{0}(E/E_{0})^{-\Gamma}$ cm$^{-2}$ s$^{-1}$ MeV$^{-1}$), \jj\ is detected with a Test Statistic (TS) value of {39 (notionally 5.9 $\sigma$)} and a spectral index of 2.39$\pm$0.10$_{stat}\pm$0.05$_{sys}$, yielding an energy flux of ({1.25} $\pm$ 0.24$_{stat}\pm$ 0.39$_{sys}$) $\times$ 10$^{-11}$ erg~cm$^{-2}$s$^{-1}$, {corresponding to a luminosity of 3.2$\times$10$^{34}$ erg~s$^{-1}$ ($d=4.6$ kpc\cite{Gaia2016})}.
No morphological extension nor spectral cutoff can be identified (see \emph{Methods}).

The GeV excess in the west is spatially coincident with the west lobe of \sss\/,
and is located at {R.A. = 287.46$\degree$$\pm$0.09$\degree$, decl.= 4.98$\degree$$\pm$0.08$\degree$}.
Assuming a power-law spectral shape, the likelihood analysis of the excess results in a TS value of 15 (notionally 3.5 $\sigma$), which is below the formal source detection threshold (TS=25, see \emph{Methods}).
This dim west excess has a spectral index of {2.30}$\pm$0.16$_{stat}\pm$0.11$_{sys}$  and an energy flux of ({0.75}$\pm$0.25$_{stat}\pm$0.41$_{sys}$) $\times$ 10$^{-11}$ erg~cm$^{-2}$s$^{-1}$.

\begin{center}
\begin{figure*}
\centering
\includegraphics[scale=0.276]{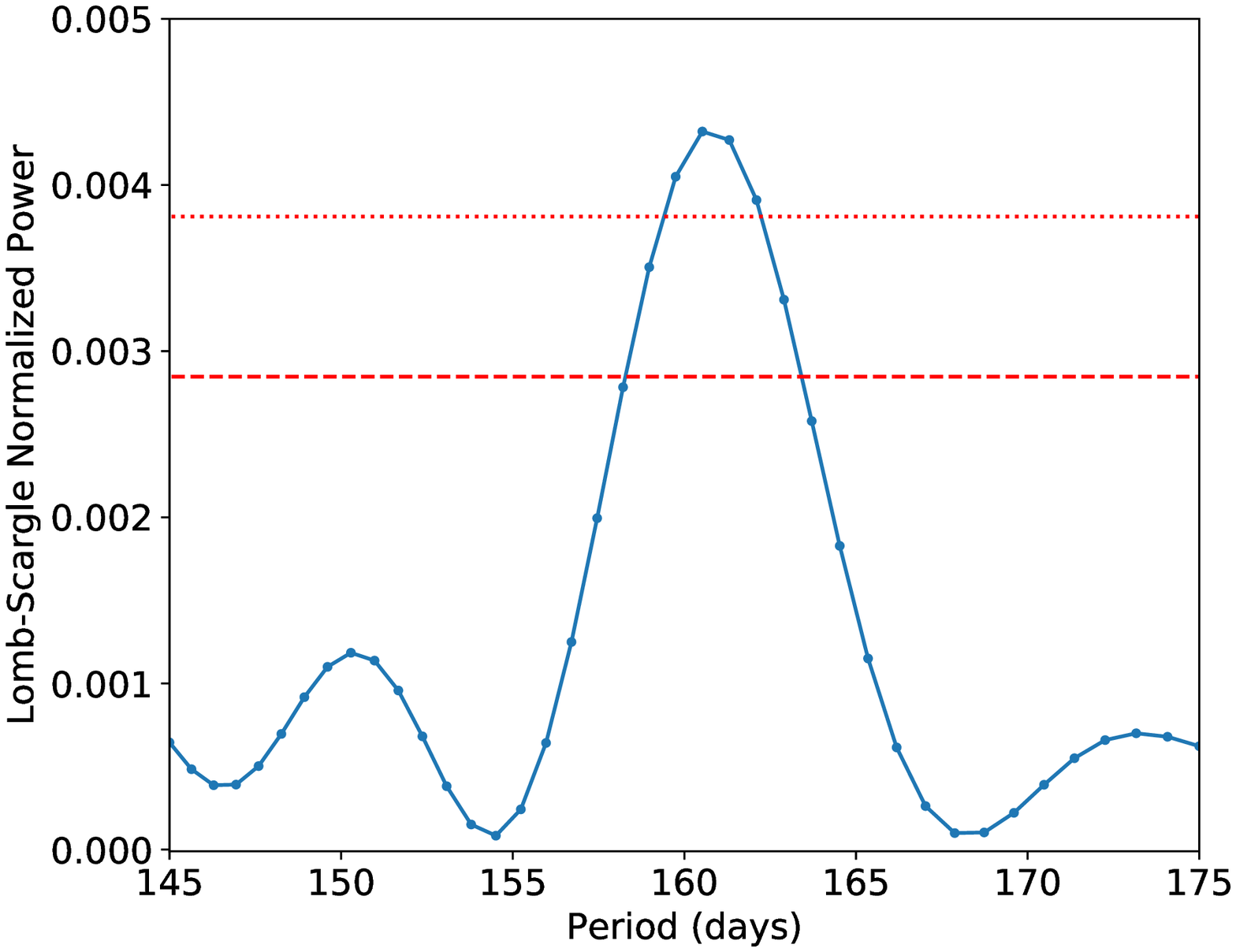}
\includegraphics[scale=0.276]{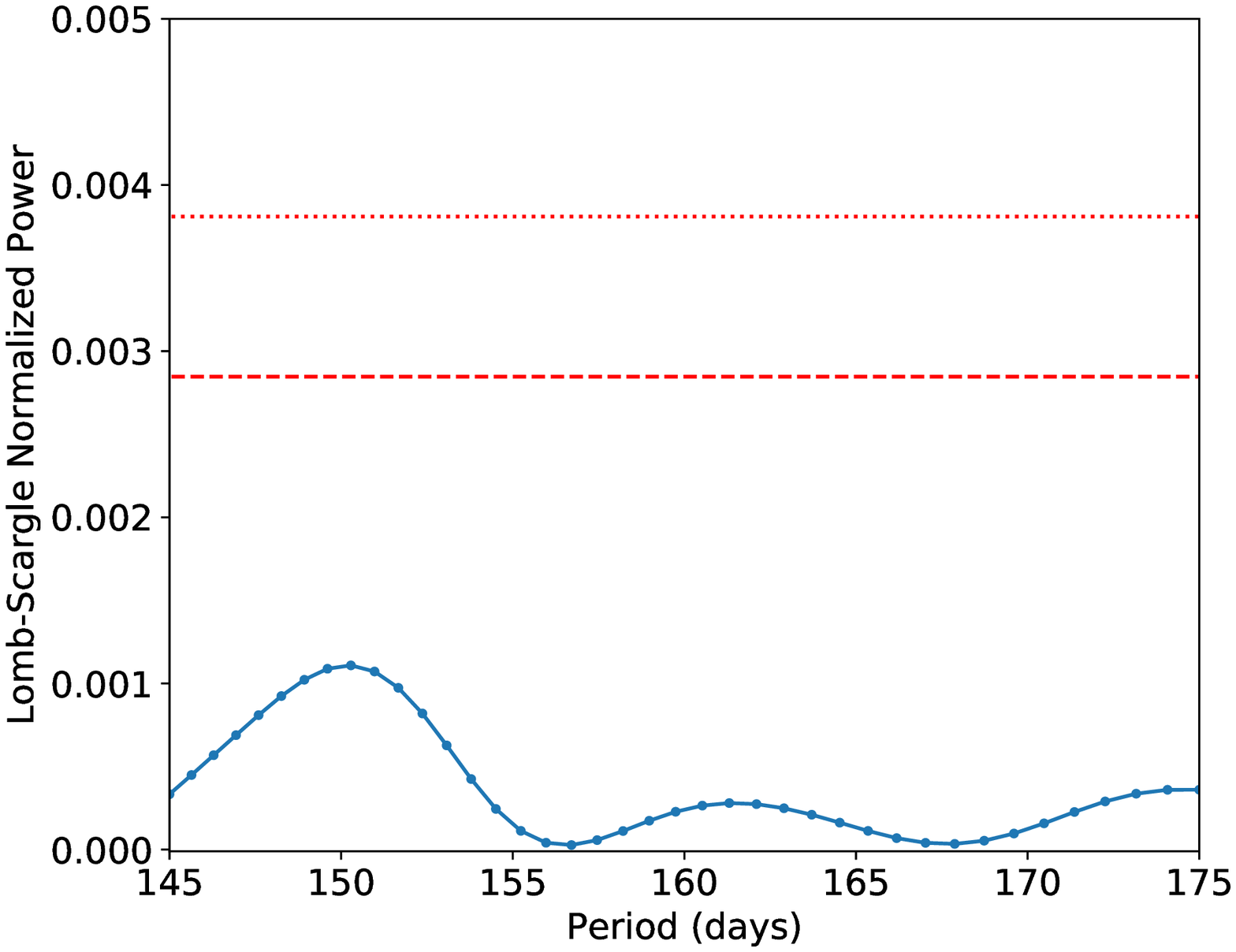}
\includegraphics[scale=0.276]{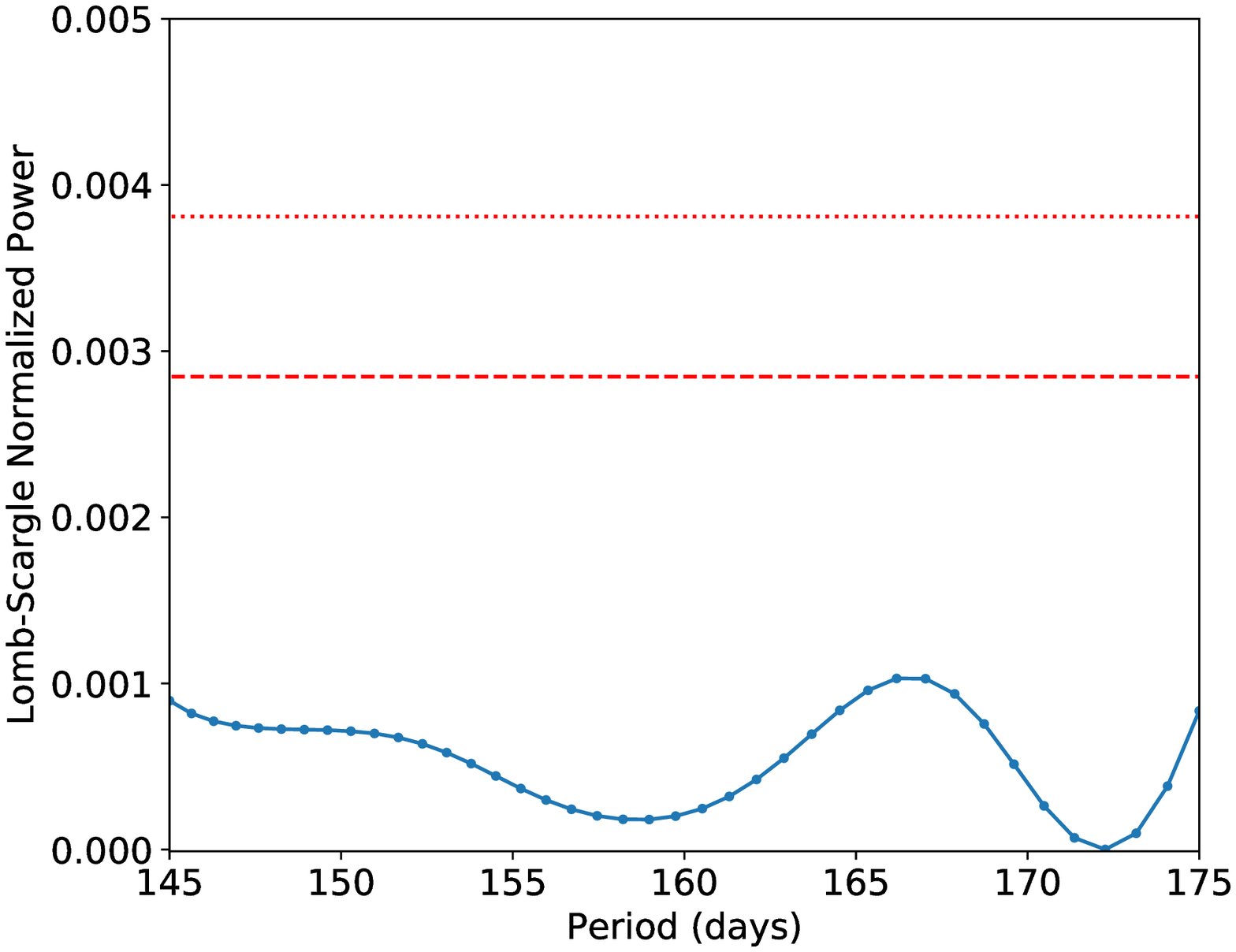}
\caption{SS 433 precession signal seen in \jj\/.
Exposure-corrected Lomb-Scargle power spectra constructed from the 1--300 GeV weighted light curve of
 \jj\/, the west excess, and \psrj\/.
The red dotted and dashed line indicates false alarm probability of 1\% and 5\% level.
Only \jj\ shows a significant hint for the detection of the precessional period, which is confirmed by likelihood analysis.
}

\label{lomb}
\end{figure*}
\end{center}

\begin{center}
\begin{figure*}[!h]
\begin{minipage}[tb]{1\textwidth}
\flushright
\includegraphics[scale=0.8]{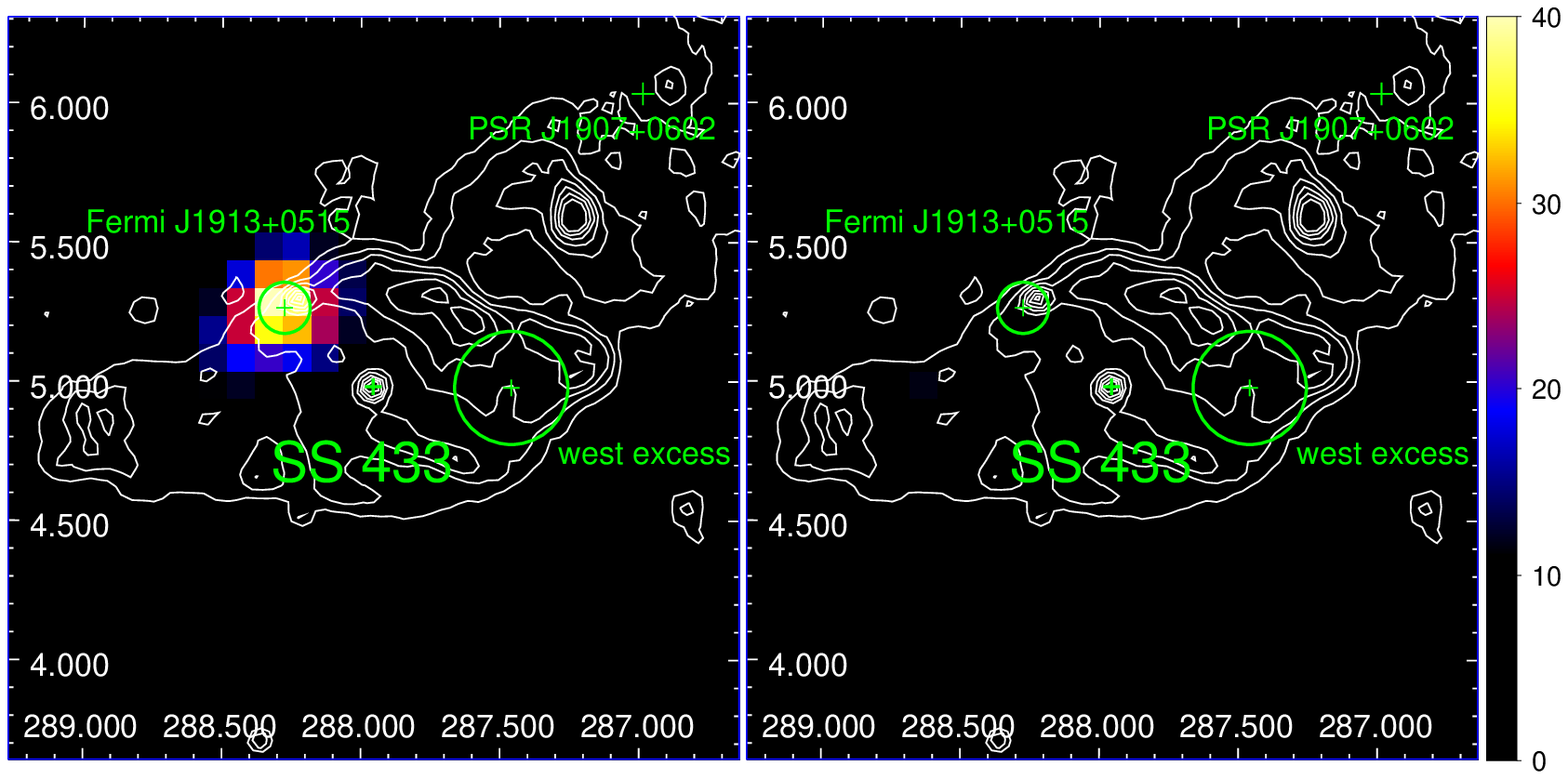}
\end{minipage}

\begin{minipage}[tb]{0.895\textwidth}
\flushright
\includegraphics[scale=0.73]{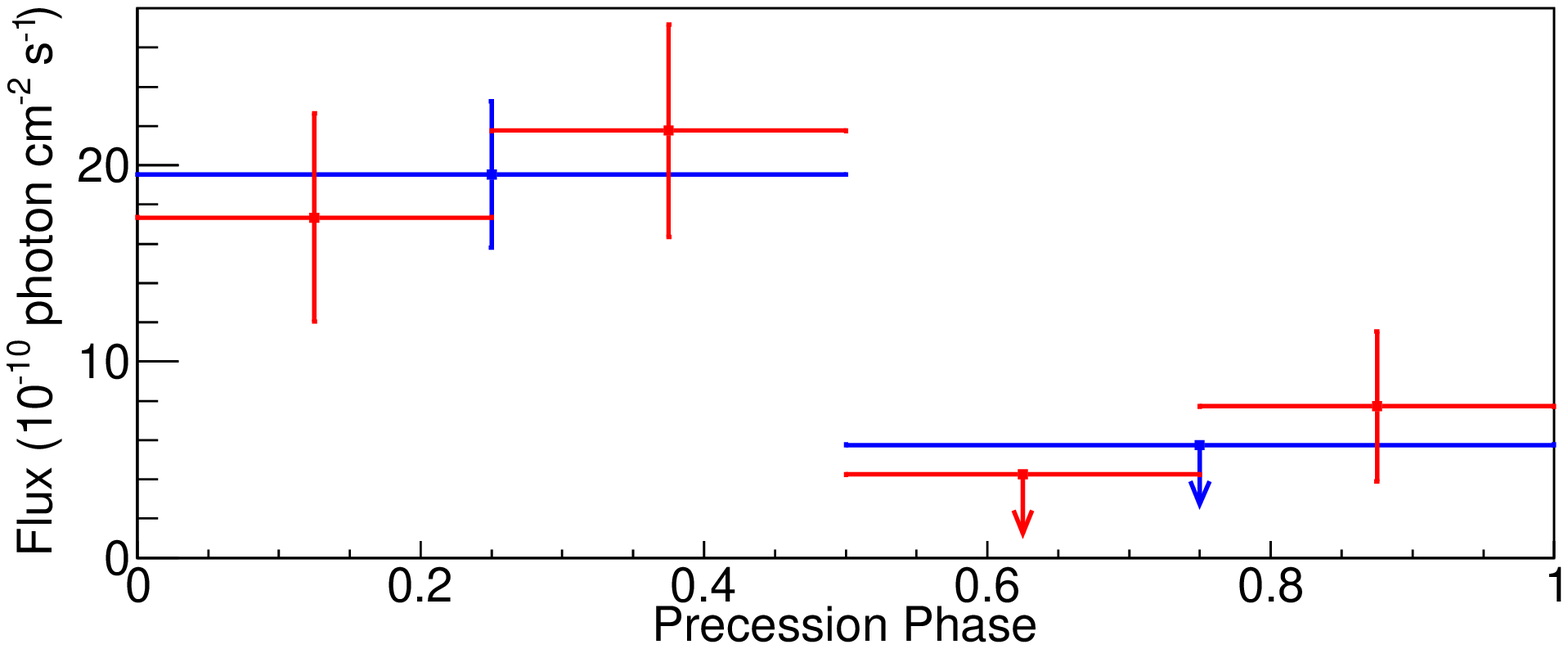}
\end{minipage}
\caption{Analysis of \emph{Fermi}-LAT data in precessional phases 0.0--0.5 (left) and 0.5--1.0 (right).
{\bf Top:} TS map of the \sss\/ region in the {1--300} GeV band for the precessional phases 0.0--0.5 (left) and 0.5--1.0 (right).
The contours and labels are as in Figure \ref{tsmap}.
{\bf Bottom:} Precessional phase light curve of \jj\/ in 1-300 GeV {with a binning of 0.5 (blue) and 0.25 (red)}.
The upper limits are at the 95\% confidence level.
}
\label{precession}
\end{figure*}
\end{center}

In an attempt to explore whether these excesses are linked to \sss\/, we produced exposure-corrected, weighted 1-day light curves above 1 GeV (see \emph{Methods}) and searched for timing signals at the orbital and precessional period.
Using Lomb-Scargle timing analysis, a hint of a periodic signal at {160.88$\pm$2.66} days is detected from \jj\ with a single-frequency significance of {3.6$\sigma$} and a false alarm probability of 3.7$\times$10$^{-3}$ (see \emph{Methods}).
This period is consistent with the jet precession period of 162.250 days (Figure \ref{lomb}).
Neither the west excess nor other sources in the vicinity show the same periodicity, and none of the sources indicates variability at the orbital period.
\jj\ itself is significantly detected, with a TS value of 31 (5.2$\sigma$) above 1 GeV.
Thus, guided by the hint of precessional variability, we carried out a likelihood analysis in two broad precession phases, 0.0--0.5 and 0.5--1.0 above 1 GeV,
adopting the \sss\ ephemeris of reference (T$_{0}$ (JD) = 2443508.4098, see \emph{Methods})\cite{Davydov2008}.
The difference is marked:
\jj\ is significantly detected in the precession phase interval 0.0--0.5 with a TS value of 39 (5.9$\sigma$)
and not detected at all in the precession phase interval 0.5--1.0, yielding a TS value of 3 (1$\sigma$, Figure \ref{precession}, top panel).
The precessional phase light curve is shown in Figure \ref{precession}, bottom panel.
Through likelihood analysis, we see that the flux in the precessional phase interval 0.0--0.5 is significantly higher than that in precession phase interval 0.5--1.0 at the 4.2$\sigma$ level.
The fluxes between the two precessional phase intervals significantly deviate from a constant at a 3.5$\sigma$ level (see \emph{Methods}).
%
%
%

%%%%%%%%%%%%%%%%%%%%%%%%%%%%%%%%%%%%%%%%%%%%%
% Discussion
%%%%%%%%%%%%%%%%%%%%%%%%%%%%%%%%%%%%%%%%%%%%%

The location of \jj\/ and the west excess nearby SS 433 reported here argues for possible physical connections.
On the one hand, the 95\% confidence level position circle of the west excess covers the X-ray excess\cite{Safi-Harb1997} and is close to the recent multi-TeV source detected by HAWC, for which a leptonic, locally-accelerated origin was energetically preferred\cite{HAWC2018}.
On the other hand, the adjacent position, the {160.88$\pm$2.66}-days timing signal and its related flux variability link \jj\ to the microquasar \sss.
However, this connection poses a significant interpretation challenge: Which is the mechanism powering the GeV emission? How is this periodic signal generated?

Detectable gamma-ray signals from SS 433 and other microquasars have been predicted in the past, even at a roughly compatible
level to the sources described here\cite{Aharonian1998,Romero2003,Reynoso2008}.
Moreover, due to the precessional period of \sss\/, a periodicity in the emission of GeV photons has also been predicted\cite{Reynoso2008}.
This signal was proposed to originate in the periodically varying --along with the precessional movement-- gamma-ray absorption due to interactions with matter (via $\gamma N$) and fields (via $\gamma \gamma$ processes) of the disk and the star\cite{Reynoso2008b}.
In this scenario,
the dominant gamma-ray production channel is hadronic, and emission must happen at the very base of the jets, at sufficiently
high ambient densities so that $pp$ interactions can proceed.
Since the same proton population would also generate copious freely-escaping neutrinos, the lack of a detected neutrino source at the position of \sss\cite{Aartsen2017} requires that the model parameters be calibrated such that they allow both a detectable signal in gamma rays and a non-detectable neutrino flux\cite{Reynoso2019}.
This is slightly complicated by the fact that the relativistic proton distribution derived in\cite{Reynoso2008} has been later revised\cite{Torres2011}, leading  to
deviations for the assumed proton fluxes for jets displaying large Lorentz factors and/or small viewing angles.
The latest H.E.S.S. and MAGIC upper limits\cite{MAGIC-HESS2018} on the central source
would require that the fractional power carried by relativistic protons in the \sss\/ jets be $\leq 10^{-5}$.
Future observations with the next generation of gamma-ray and neutrino detectors will assess this possibility  further.
However,
due to the significant positional offset between the predicted and the detected source  ($\sim$35 pc away from the central source, at a distance of $\sim$4.6 kpc\cite{Gaia2016}),
we can already conclude that the periodicity reported for \jj\ cannot be related to such gamma-ray absorption.

Coincident in position with \jj\, and at the consistent distance as SS 433, there is a gas enhancement beyond the diffuse average (Figure \ref{tsmap}, bottom panel).
The gas excess is located within a projected region of $R_c \sim 20$ pc, with a mass that can reach up to
$M \sim 25000$ M$_\odot$ (see { \emph{Methods}}). Assuming a spherical region of radius $R_c$,
the average density is $n\sim22$ cm$^{-3}$, although can be lower if the mass is more extended perpendicular to the plane of the sky.

Direct periodic illumination of {such region} by the eastern jet seems unlikely.
On the one hand, the coherence of the radio jet appears to be sustained on the arcsecond scale only\cite{Blundell2004}.
Simulations confirm that the jet loses the helical morphology after a few precession cycles, due to the interaction with the surrounding medium\cite{Monceau2015}.
On the other hand,   \jj\ is not located within the extrapolated jet cone.

The interaction of protons accelerated in the central region of microquasars or at the jet termination in neighboring  clouds has been studied in the past\cite{Bosch2005}.
In such scenarios,
protons diffuse from their injection point
and produce hadronic gamma-rays when finding appropriate targets.
The average level of gamma-ray flux we measure from \jj\ can be accommodated in this setting.
This scenario, however, can hardly explain the periodicity.
Even assuming a periodic, impulsive injection of cosmic rays containing most of the jet energy released in a single period,
{these injections} would not be energetically relevant individually, providing a cosmic-ray density subdominant to the Galactic sea.
The \emph{Methods} section gives further details on these considerations.

An alternative possibility for proton injection could be provided by the relativistic equatorial outflow
recently characterized by \emph{NuSTAR}\cite{Middleton2018}:
This outflow has a more favorable geometry with respect to the gas enhancement.
The line-of-sight outflow velocity is 0.14--0.29c (potentially higher if we are not viewing along the direction of the outflow), and it even
exceeds the velocities seen in the approaching jet ejecta at any precessional phase\cite{Middleton2018}.
Energetically, the outflow is as powerful as that of the jet
and is believed to precess in solidarity with the jet and the accretion disk.
The screening of the central source by these outflows
would explain why \sss\ is not as X-ray bright as the 10$^{-4}$ M$_\odot$ yr$^{-1}$ accretion rate would indicate\cite{Middleton2018}.

However, and similarly to what was noted above, in order to have proton interactions associated with the precessional periodicity, protons should arrive at {a} cloud periodically at a sufficient rate to produce the gamma-ray emission level seen.
Anisotropic diffusion\cite{Nava2013} or streaming of cosmic rays along a flux tube could perhaps help:
In anisotropic diffusion, the cosmic-ray density along the tube, at a generic distance $R_t$ from the source,  would be
proportional to\cite{Nava2013} $n \propto (R_{t} R_{src}^{2} )^{-1}$ instead of $R_{t}^{-3}$ as in the isotropic case (where $R_{src}$
is the typical scale for the accelerating region, which for SS 433 can be much smaller than 1 pc).
If the gas enhancement and \sss\ are periodically connected via a magnetic flux tube within the W50 nebula,
or if the very injection at the base of the magnetic tube is periodic,
a sufficient number of protons could arrive and interact at the cloud in each period.
In this scenario,
most of the accelerated protons in one period, i.e., a reservoir up to $4\times 10^{45}\rm erg$, considering a kinetic luminosity of $3\times 10^{39}\rm erg$ s$^{-1}$ and 10\% of it converting to cosmic rays, might be assumed to reach the region of interest, which eases the energetic requirements.
Maintaining periodic coherence, however, requires cosmic rays being funnelled into the gas enhancement
and interacting in a small clump or cusp of density (in what otherwise is a relatively low-density gas enhancement, see \emph{Methods}). Thus, this is a priori unlikely too,
although further studies are required to rule it out.
The periodic variability  is  intriguing,
and difficult to reconcile with our current understanding of the source environment under common lore interpretations.

SS 433 continues to amaze observers at all frequencies and theoreticians alike, and is certain to provide a testbed for our ideas on cosmic-ray production
and propagation near microquasars for years to come.

\textbf{Acknowledgments} \\

The \textit{Fermi} LAT Collaboration acknowledges generous ongoing support
from a number of agencies and institutes that have supported both the
development and the operation of the LAT as well as scientific data analysis.
These include the National Aeronautics and Space Administration and the
Department of Energy in the United States, the Commissariat \`a l'Energie Atomique
and the Centre National de la Recherche Scientifique / Institut National de Physique
Nucl\'eaire et de Physique des Particules in France, the Agenzia Spaziale Italiana
and the Istituto Nazionale di Fisica Nucleare in Italy, the Ministry of Education,
Culture, Sports, Science and Technology (MEXT), High Energy Accelerator Research
Organization (KEK) and Japan Aerospace Exploration Agency (JAXA) in Japan, and
the K.~A.~Wallenberg Foundation, the Swedish Research Council and the
Swedish National Space Board in Sweden. Additional support for science analysis during the operations phase is gratefully acknowledged from the Istituto Nazionale di Astrofisica in Italy and the Centre National d'\'Etudes Spatiales in France. {This work performed in part under DOE Contract DE-AC02-76SF00515.}

This publication utilizes data from Galactic ALFA HI (GALFA HI) survey data set obtained with the Arecibo L-band Feed Array (ALFA) on the Arecibo 305m telescope. The Arecibo Observatory is operated by SRI International under a cooperative agreement with the National Science Foundation (AST-1100968), and in alliance with Ana G. M$\acute{e}$ndez-Universidad Metropolitana, and the Universities Space Research Association. The GALFA HI surveys have been funded by the NSF through grants to Columbia University, the University of Wisconsin, and the University of California.

J. L. acknowledges the support from the Alexander von Humboldt Foundation and the National Natural Science Foundation of China via NSFC-11673013, NSFC-11733009.
The work of D. F. T. has been supported by the grants PGC2018-095512-B-I00, SGR2017-1383, and AYA2017- 92402-EXP.
J. L.  and D. F. T. acknowledge the fruitful discussions with the international team on ``Understanding and unifying the gamma rays emitting scenarios in high mass and low mass X-ray binaries" of the ISSI (International Space Science Institute), Beijing,
as well as the support of the PHAROS COST Action (CA16214).
E.O.W acknowledges the support from the Alexander von Humboldt Foundation.
Work at NRL is supported by NASA.
We thank Dr. Rolf B$\ddot{u}$hler, Dr. Fabio Acero, Dr. Jean Ballet, Dr. Philippe Bruel, Dr. David J. Thompson, Dr. Seth Digel and Dr. Gu\dh laugur J$\acute{o}$hannesson for their insightful comments and helpful suggestions.
We thank Dr. Peng Zhang for the help with Weighted Wavelet Z-transform.

%%%%%%%%%%%%%%%%%%%%%%%%%%%%%%%%%%%
%%%%%%%%%%%%%%%%%%%%%%%%%%%%%%%%%%

\label{results}
%%%%%%%%%%%%%%%%%%%%%%%%%%%%%%%%%%%
%%%%%%%%%%%%%%%%%%%%%%%%%%%%%%%%%%

%%%%%%%%%%%%%%%%%%%%%%%%%%%%%%
%%%%%%%%%%%%%%%%%%%%%%%%%%%%%%%%%%

\clearpage

%%%%%%%%%%%%%%%%%%%%%%%%%%%%%%%%%%%%%%%%%%%%%
\textbf{Methods}
%%%%%%%%%%%%%%%%%%%%%%%%%%%%%%%%%%%%%%%%%%%%%

%%%%%%%%%%%%%%%%%%%%%%%%%%%%%%%%%%%%%%%%%%%%%
\textbf{Details on the {\it Fermi}-LAT data analysis}
%%%%%%%%%%%%%%%%%%%%%%%%%%%%%%%%%%%%%%%%%%%%%

The analysis shown in this paper uses 10.5 years of \emph{Fermi}-LAT\cite{Atwood2009} data, from 2008 August 4 (MJD 54682) to 2019 January 28 (MJD 58511).
We have considered all events with reconstructed energies between 100 MeV--300 GeV and positions within a circular region of interest (ROI) of 15$\degree$ radius centered on \psrj.
We selected photons of the ``Pass 8'' event class, using \emph{Fermi} Science Tools\cite{fermi-tools} 11-07-00 release.

We have used the ``P8R3 V2 Source'' instrument response functions (IRFs),
adopting a zenith angle threshold of $<$ 90$\degree$ to reject contaminating gamma rays from the Earth's limb.
A spectral-spatial model was constructed from the \textit{Fermi} Large Area Telescope Fourth Source Catalog (4FGL)\cite{4FGL}.
Both Galactic (``gll\_iem\_v07.fits") as well as isotropic diffuse emission components  (``iso\_P8R3\_SOURCE\_V2\_v1.txt"\cite{bkg}) and known gamma-ray sources within $20\degree$ of \psrj\/ were included.

The spectral parameters of the sources within 4$\degree $ of our target were left free while those of other (farther) sources included were fixed at the 4FGL values.
The spectral analysis was performed using a binned maximum likelihood fit (spatial bin size 0.1 degree, AIT projection and 30 logarithmically spaced bins in the 0.1--300 GeV range) with the Science Tool \emph{gtlike}.
The significance of the sources were evaluated by the Test Statistic (TS).
This statistic is defined as TS=$-2 \ln (L_{max, 0}/L_{max, 1})$, where $L_{max, 0}$ is the maximum likelihood value for a model in which the source studied is removed (the ``null hypothesis"), and $L_{max, 1}$ is the corresponding maximum likelihood value with this source being incorporated.
The larger the value of TS, the less likely it is that the null hypothesis (no source) is correct, and that instead a significant gamma-ray excess lies at the tested position.
For nested models with nondegenerate parameters, $\sqrt{TS}$ is approximately equal to the detection significance in $\sigma$ of a given source.
A TS of 25 is adopted as the detection threshold in this paper, as is similarly done in the {\it Fermi}-LAT source catalogs\cite{Acero2015}.

The extension significance was defined as {TS$_{ext}$=$-2\ln (L_{point}/L_{ext}$)}, where $L_{ext}$ and $L_{point}$ are
the \textit{gtlike} global likelihood of the extended source hypotheses and the point source, respectively.
A threshold for claiming the source to be spatially extended is set as TS$_{ext}>$16, {which corresponds} to a significance of $\sim$ 4$\sigma$.
The \textit{Fermipy} python package\cite{wood2017} was used to produce the TS maps and source localizations in this paper.
Energy dispersion correction has been applied in the analysis.

The systematic errors have been estimated following standard procedures, i.e., by repeating the analysis using modified IRFs\cite{Ackermann2012} that bracket the effective area\cite{Aeff}, and artificially changing the normalization of the Galactic diffuse model  by $\pm$ 6\%\cite{Abdo2013}.
{The {latter} dominates the systematic errors.}
{In this paper, the} first (second) uncertainty shown corresponds to the statistical (systematic) error.

%%%%%%%%%%%%%%%%%%%%%%%%%%%%%%%%%%%%%%%%%%%
\textbf{Pulsar contamination and the need of gating}
%%%%%%%%%%%%%%%%%%%%%%%%%%%%%%%%%%%%%%%%%%%%%

\begin{center}
\begin{figure*}
\centering
\begin{minipage}[tb]{0.48\textwidth}
\centering
\includegraphics[scale=0.405]{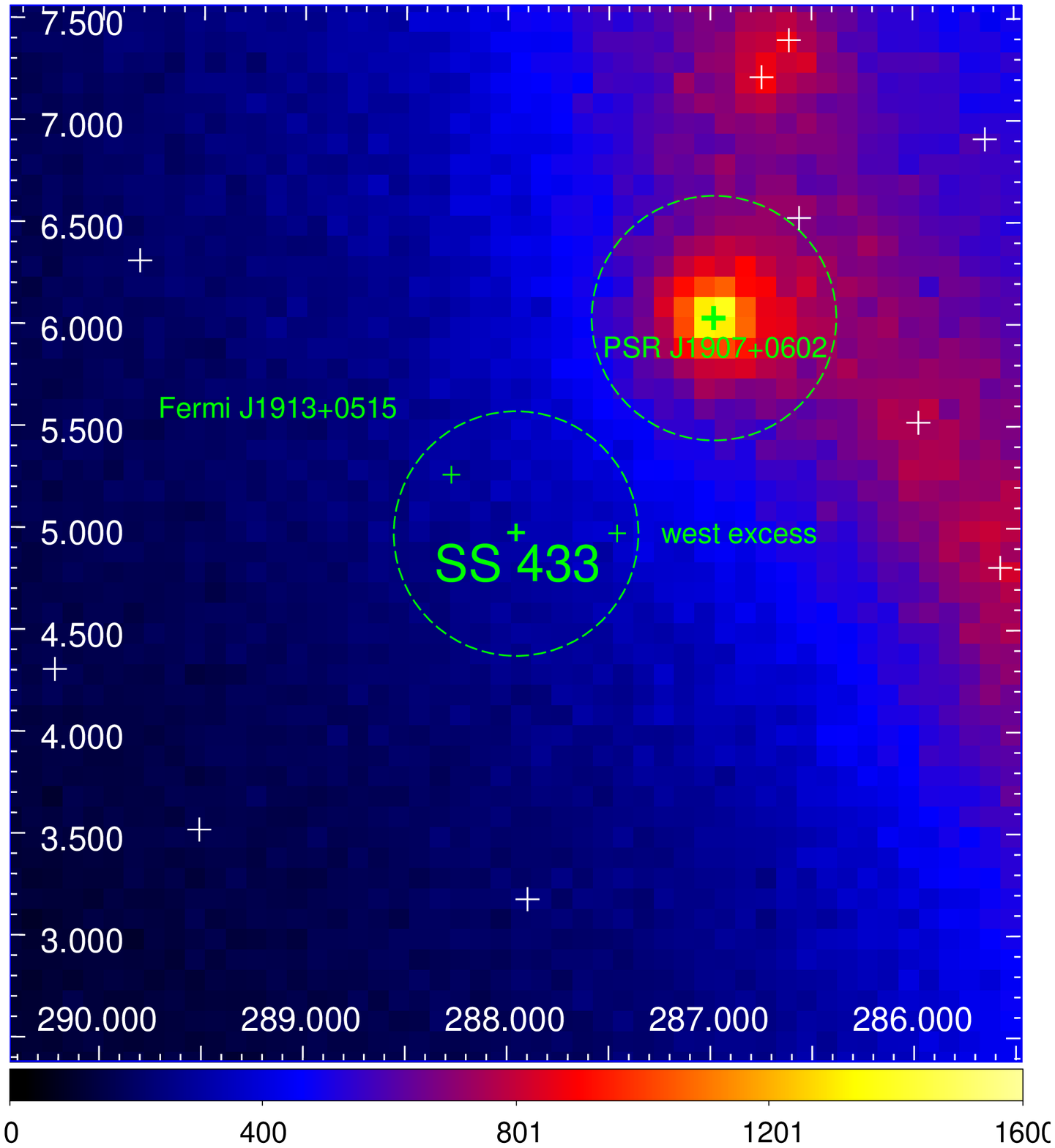}
\end{minipage}
\begin{minipage}[tb]{0.48\textwidth}
\centering
\includegraphics[scale=0.36]{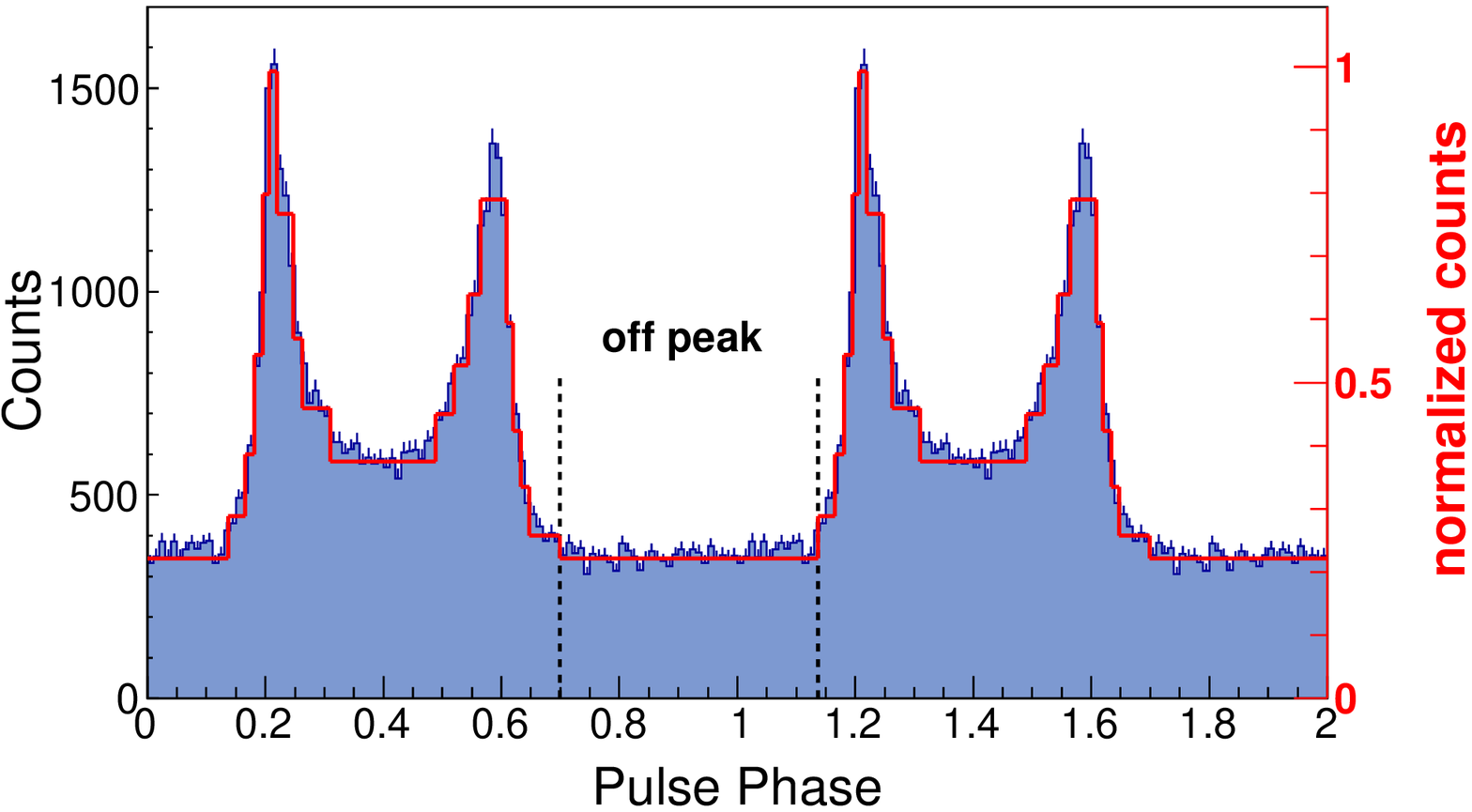}
\includegraphics[scale=0.36]{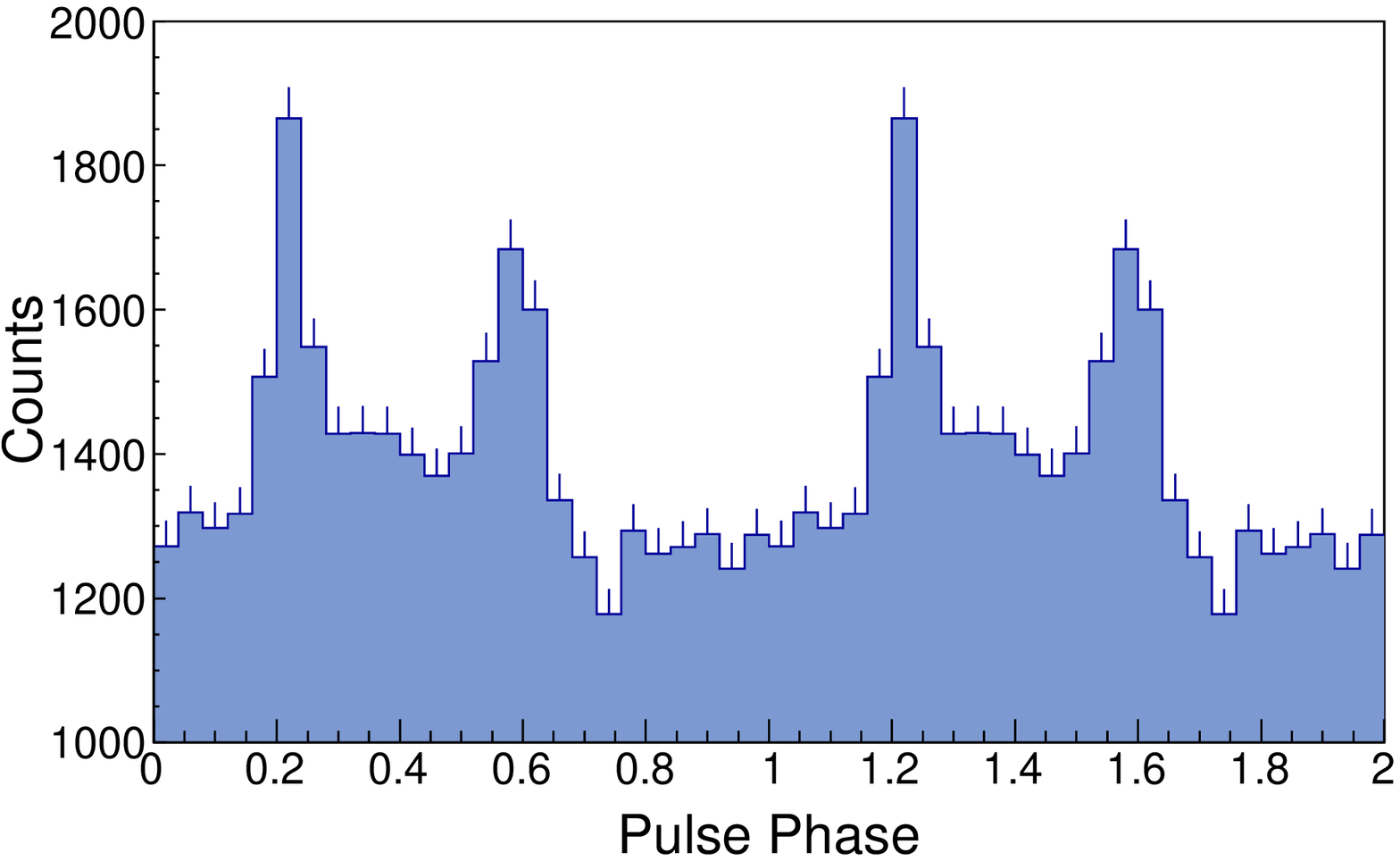}
\end{minipage}
\caption{Gamma-ray pulsar \psrj\/'s contamination on SS 433 region.
{\bf Left:} 100 MeV -- 300 GeV counts map of the \emph{Fermi}-LAT field of the SS 433 region. The microquasar itself is noted with a bold cross.
The fitted position of \jj\ and west excess are shown with green crosses.
The X- and Y-axes are R.A. and decl. referenced at J2000.
The regions used to produce pulse profiles are shown with dotted circles.
{\bf Right-top:} Folded pulse profile of \psrj\/ above 300 MeV with an ROI of 0.6$\degree$.
Two rotational pulse periods are shown, with a resolution of 100 phase bins per period.
The Bayesian block decomposition is shown by red lines.
The off-peak interval ($\phi$=0.697--1.136) is defined by black dotted lines.
{\bf Right bottom:} Folded pulse profile of the photons centered on \sss\/ with a radius of 0.6$\degree$ above 100 MeV, using the ephemeris of \psrj\/.
Two rotational pulse periods are shown, with a resolution of 25 phase bins per period.
}

\label{cmap}
\end{figure*}
\end{center}

\psrj\/ is a bright gamma-ray pulsar located only 1.4$\degree$ away from SS 433.
The photons from \psrj\/ dominate the emission of the \sss\/ region (Figure \ref{cmap}), up to the point that  \jj\/ is not visible in the counts map.
To produce the pulse profile, we selected photons from \psrj\/ above 300 MeV within a radius of 0.6$\degree$, selections which maximized the H-test statistic\cite{deJager1989,deJager2010}.

We have assigned pulsar rotational phases for each gamma-ray photon that passed the selection criteria, applying an updated ephemeris for \psrj\/ and using \emph{Tempo2}\cite{Hobbs2006} with the {\it Fermi} plug-in\cite{Ray2011}.
The pulsation from \psrj\/ is significantly detected with an H-Test value of 14948 (m=20).
 Its pulse profile is shown in Figure \ref{cmap}, right panel.

To check for the level of contamination at the \sss\/ region that may be produced by \psrj, we extracted photons above 100 MeV within a radius of 0.6$\degree$ centered on \sss\/ (dashed circle in Figure \ref{cmap}, left panel), thus covering both regions of interest for this work, \jj\ and the west excess.
Pulsar rotational phases for each gamma-ray photon were calculated at that position using the ephemeris of the pulsar \psrj.
The folded profile is shown in Figure \ref{cmap}, right panel.
It shows that the pulsation of \psrj\ is significantly recovered at the position of interest, with an H-Test value of 418 (m=11, above 8 $\sigma$).
This exercise demonstrates that a non-gated analysis of the gamma-ray photons at the position of \sss\/ is severely contaminated by
 \psrj.

Another way to confirm the contamination level is to analyze {\it gtlike} results for this dataset.
The likelihood analysis of the \psrj\/ in {100 MeV-300 GeV} yields a TS value of 22142, while for \jj\/ it yields a TS value of 54.
{We also produced model counts map of \psrj\/ and \jj\/ with \emph{gtmodel} using the likelihood analysis.} %spectral-spatial model derived  are shown in Figure \ref{model}.
On the position of \jj\/, 24 photons are expected to come from \jj\/.
In turn, 25 photons are expected to come from \psrj\/ at the same position of interest.
%

%%%%%%%%%%%%%%%%%%%%%%%%%%%%%%%%%%%%%%%%%%%%%
\textbf{Pulsar gating}
%%%%%%%%%%%%%%%%%%%%%%%%%%%%%%%%%%%%%%%%%%%%%

To minimize the contamination from the nearby pulsar, we carried out our data analysis during the off-peak phases of \psrj\/, following similar analyses\cite{abdo2010,li2017}.
To define the off-peak interval, we divided the pulsed light curve into cells using the Bayesian Blocks algorithm described in \cite{Abdo2013} (details can be found in \cite{jackson2015} and \cite{scargle2013}).
The off-peak interval is then defined as $\phi$=0.0$-$0.136 and 0.697$-$1.0 (see Figure \ref{cmap}, right panel).
Correspondingly, the prefactor parameters of all sources were scaled by 0.439 in the pulsar off-peak interval analysis, which is the width of the off-peak interval.
The off-peak emission of \psrj\ was modelled with a power-law.

%%%%%%%%%%%%%%%%%%%%%%%%%%%%%%%%%%%%%%%%%%%%%
\textbf{Off-peak analysis}
%%%%%%%%%%%%%%%%%%%%%%%%%%%%%%%%%%%%%%%%%%%%%

\begin{center}
\begin{figure*}
\centering
\includegraphics[scale=0.405]{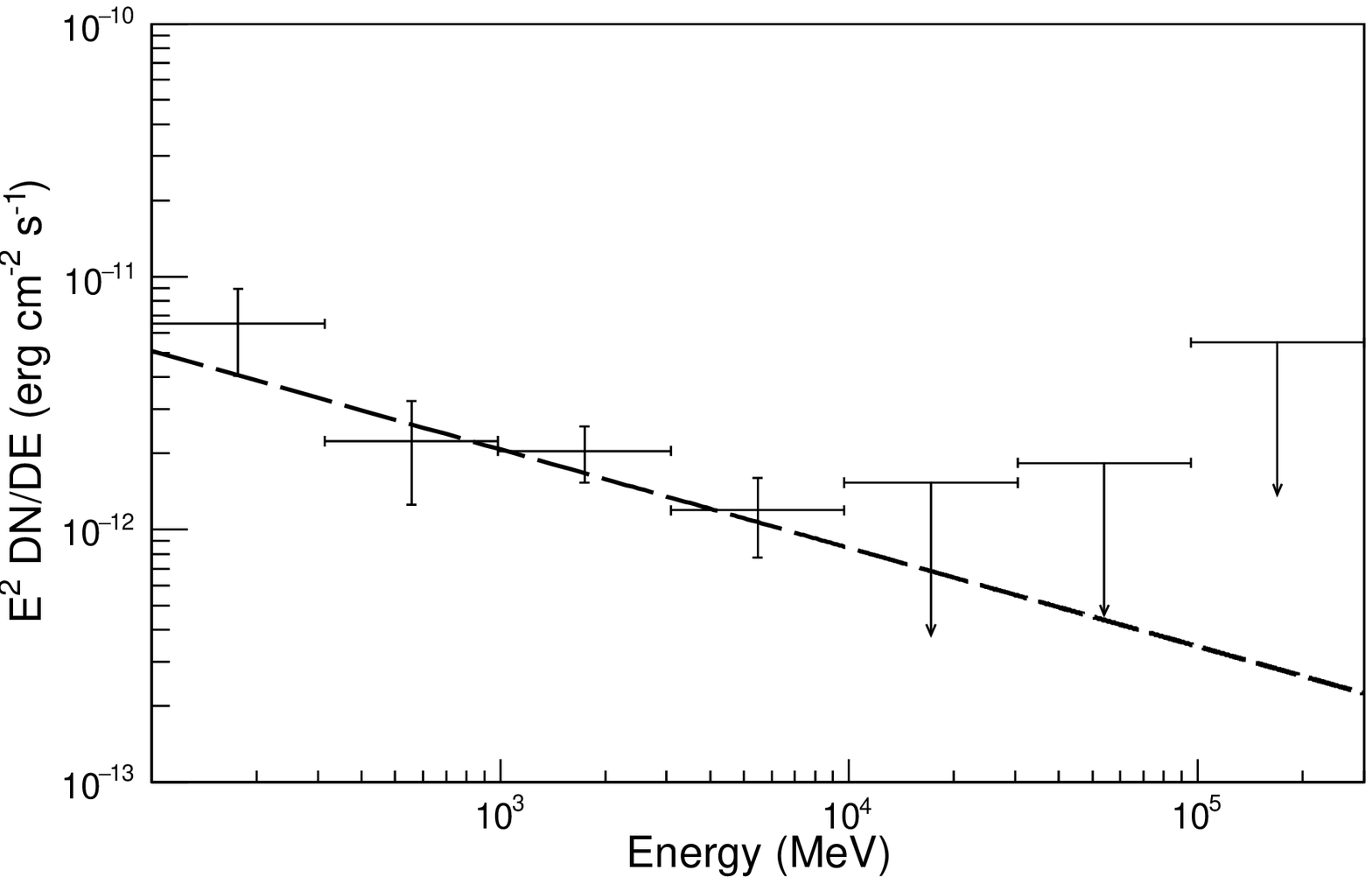}
\includegraphics[scale=0.405]{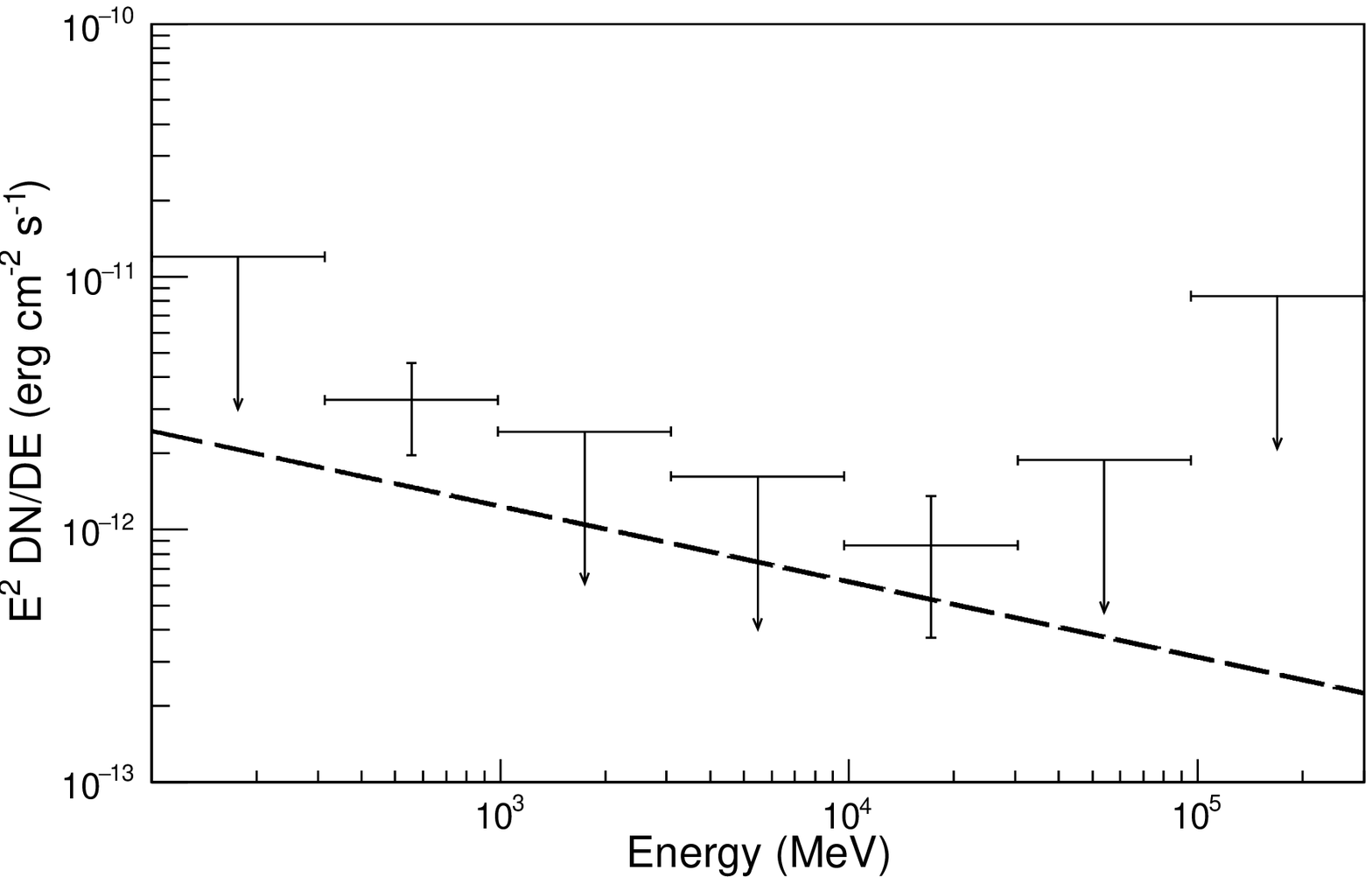}
\caption{\emph{Fermi}-LAT spectra of \jj\ (left) and the west excess (right). The maximum likelihood model (power law) fitted with \emph{gtlike} is shown with a dashed line.}
\label{SED}
\end{figure*}
\end{center}

During the off-peak of \psrj\/, and assuming a power-law spectral shape, \jj\ is detected with a TS value of 39 ($\sim 5.9\sigma$) and a spectral index of 2.39$\pm$0.10$_{stat}\pm$0.05$_{sys}$, yielding an energy flux of ({1.25} $\pm$ 0.24$_{stat}\pm$ 0.39$_{sys}$) $\times$ 10$^{-11}$ erg~cm$^{-2}$s$^{-1}$.
We also modelled \jj\ with a power law with an exponential cutoff ($dN/dE=N_{0}(E/E_{0})^{-\Gamma}$exp$(-E/E_{0}) $ cm$^{-2}$ s$^{-1}$ MeV$^{-1}$).
However, the $\Delta$TS ($\Delta$TS=$-2 \ln (L_{PL}/L_{CPL})$, where $L_{CPL}$ and $L_{PL}$ are the maximum likelihood values for power-law models {with and without} a cutoff) between the two models is {less than 9}, which indicates that a cutoff is not significantly preferred.
The extension of \jj\ was searched using \emph{Fermipy} but no extension was found to be significant.
For a further test, using \emph{gtlike} we modelled \jj\ as a disk of different radius (from 0.1 to 0.5 degree with a step of 0.1 degree), all yielding a TS$_{ext}$ < 16.
No significant extension is detected.
Assuming a power-law spectral shape, the likelihood analysis of the west excess resulted in a TS value of 15, a spectral index of {2.30}$\pm$0.16$_{stat}\pm$0.11 $_{sys}$, and an energy flux of ({0.75}$\pm$0.25$_{stat}\pm$0.41$_{sys}$) $\times$ 10$^{-11}$ erg~cm$^{-2}$s$^{-1}$.
Since this excess is not significantly detected (TS below the source detection threshold of 25),
we do not consider its possible extension or the existence of a spectral cutoff.
The TS map of \sss\/ region is shown in Figure \ref{tsmap}, top panel, together with the radio morphology from the Effelsberg 11 cm survey\protect\cite{Reich1990}.
The GeV spectral energy distribution (SED) of \jj\ and the west excess are shown in Figure \ref{SED}.

Out of academic interest, and to compare with the former as well as with previous publications, we also carried out an analysis of the region without pulsar gating.
As expected (see the discussion above regarding photon contamination)
an extended/softer GeV excess is spatially associated with SS 433/W50 in this situation,
similar to the results reported in \cite{Sun2019}.
\jj\ and the west excess can not be distinguished anymore.
\jj\ would have a softer power-law index of 2.53$\pm$0.03.
All of these result from the inclusion of photons that are not coming from the regions of interest, but from the contaminating pulsar, as demonstrated before.

%%%%%%%%%%%%%%%%%%%%%%%%%%%%%%%%%%%%%%%%%%%%%
\textbf{Differences between the use of FL8Y and 4FGL}
%%%%%%%%%%%%%%%%%%%%%%%%%%%%%%%%%%%%%%%%%%%%%

We note that \jj\/ is associated with FL8Y J1913.3+0515 but that no 4FGL source is reported at this position.
This is a result of a change in the diffuse model used in the 4FGL.
Indeed, the FL8Y list has 5523 sources\cite{FL8Y} while the 4FGL catalog has 5065 sources\cite{4FGL}.
The two catalogs used the same amount of data and software, but different interstellar emission model (gll\_iem\_v06 for FL8Y and gll\_iem\_v07 for 4FGL), also
different energy range (100MeV--1TeV for FL8Y and 50MeV--1TeV for 4FGL) and a different threshold for using a
curved spectral shape (TS$_{curve}$>16 for FL8Y and TS$_{curve}$>9 for 4FGL).
The different interstellar emission model (higher at lower energies in the new model) used in each catalog is the main reason for the disappearance of sources.
As stated in the 4FGL paper, changing the Galactic diffuse emission model from gll\_iem\_v06 to gll\_iem\_v07, even without changing the analysis or the data, the number of sources detected decreased by 10\%.

Indeed, we carried out an analysis of 8 years P8R3 data in 50 MeV--1 TeV,
which is the same time period
and energy range used for the 4FGL.
Using the 4FGL and assuming a power-law model, \jj\ is not significantly detected, with TS=19, which is consistent with the absence
of a corresponding source in this list.
However, using the 4FGL with the previous version of Galactic diffuse emission model (gll\_iem\_v06), Fermi J1913+0515 is again significantly detected with TS=36.

We have also checked our off-peak analysis presented in this paper using the FL8Y catalog.
Both \jj\ and the west excess are significantly detected with TS=73 and TS=34, respectively.
Finally, we recall that above 1 GeV, the difference of results using FL8Y and 4FGL is minor and all the results included in this paper are consistent in both analysis.

%%%%%%%%%%%%%%%%%%%%%%%%%%%%%%%%%%%%%%%%%%%%%
\textbf{Weighted light curve}
%%%%%%%%%%%%%%%%%%%%%%%%%%%%%%%%%%%%%%%%%%%%%

Adopting the best-fit spectral-spatial model derived from the precessional phase-averaged analysis, we selected photons within a 3$\degree$-radius of \sss\/ and calculated the probability that each event originated from \jj\/ using \textit{gtsrcprob}.
For a better Point Spread Function (PSF) and less contamination from background sources, we only considered events above 1 GeV.
Binning into 1-day intervals and correcting for the instrument exposure produced a light curve.
We searched for the precessional periodic signal in the light curve between 145 and 175 days using the Lomb-Scargle periodogram method\cite{Lomb1976,Scargle1982}.
Power spectra around the 162.250 days precession period were generated for the \jj\/ exposure-corrected and exposure-uncorrected light curves {using the Python packages \emph{astropy} and \emph{PyAstronomy}}.
No significant periodic signal was discovered in the uncorrected light curve.
However, after the exposure correction is applied, a {160.88$\pm$2.66} days period is detected {with a single frequency significance of }{3.6}$\sigma$, consistent with the 162.250 days jet precession period.
The single frequency significance was estimated using Python package \emph{PyAstronomy}.
``Standard'' normalization method in Python package \emph{astropy} was used.
{To calculate the false alarm probability and estimate how likely it is for the timing signal we detected to have an origin in noise, we implemented a bootstrap method.
To construct the simulated light curve, we keep the temporal coordinates the same as the actual light curve and assumed a Gaussian white noise for the flux.
We computed Lomb-Scargle periodograms on 10$^{5}$ resampled, simulated light curves and derived the false alarm probability of the
detected timing signal at our period of interest.
All the sampled periods in the Lomb-Scargle periodogram have been considered in the bootstrap.
Our results are shown in Figure 2 of the main text.}

The Lomb-Scargle timing analysis has also been checked with different binning (e.g. 5 days, 7 days, 10 days) of the weighted light curve.
The results are all consistent and do not depend on the weighted light curve binning.
In addition, we note that
the Fourier period resolution ($P^{2}/2T$,  where $P$ is the trial period and $T$ is the observation interval) of our 10.5 years light curve at $\sim$160 days is $\sim$3.4 days.
To infer the best period, the Fourier period resolution is usually oversampled by a factor of several, see e.g.,\cite{Israel2019}.
To obtain the results above, we oversampled it by a factor of  $\sim$4, leading to a period resolution of $\sim$0.8 days around 160 days.
However,
the evidence for the periodic signal at 160.88$\pm$2.66 days do not depend on the period resolution adopted in the Lomb-Scargle periodogram:
Other period resolutions around 160 days (e.g. $\sim$1.0, $\sim$1.6 and $\sim$ 3.4 days -- i.e., with no oversampling) have been tested and all lead to consistent results.

Using the same light curve and a similar method, a search for the 13.082 days orbital periodic signal was carried out between 10 and 20 days, leading to no detections.

%%%%%%%%%%%%%%%%%%%%%%%%%%%%%%%%%%%%%%%%%%%%%
\textbf{Likelihood analysis of the flux variation between precession phase 0.0--0.5 and 0.5--1.0}
%%%%%%%%%%%%%%%%%%%%%%%%%%%%%%%%%%%%%%%%%%%%%

We adopted the precession period of 162.250 days from \cite{Davydov2008}.
The precessional phase zero (T$_{0}$) is set to the time of the largest separation of the moving emission lines in SS 433, JD 2443508.4098 (referred to as T$_{3}$ in \cite{Davydov2008}).
To estimate the significance of the flux variation between the precessional phase 0.0--0.5 and 0.5--1.0 described in the main text, we employed a likelihood analysis in 1--300 GeV.
The two data sets of precessional phase 0.0--0.5 and 0.5--1.0 are jointly fitted using summed likelihood analysis.
An additional co-spatial source with \jj\ is added for precession phase 0.0-0.5 to model any flux excess from that in the precession phase 0.5-1.0.
With spectral index fixed at the value of \jj\/ independently derived in precessional phase 0.0--0.5, the co-spatial source yields a TS value of 18 ($\sim$ 4.2$\sigma$), and further demonstrates the significance of the flux difference between the two precessional phase bins.
To further explore the trend of flux modulation, we show the precessional phase light curve of \jj\/ in 1-300 GeV with a binning of 0.25 (Figure \ref{precession}).

\begin{center}
\begin{figure*}
\centering
\includegraphics[scale=0.5]{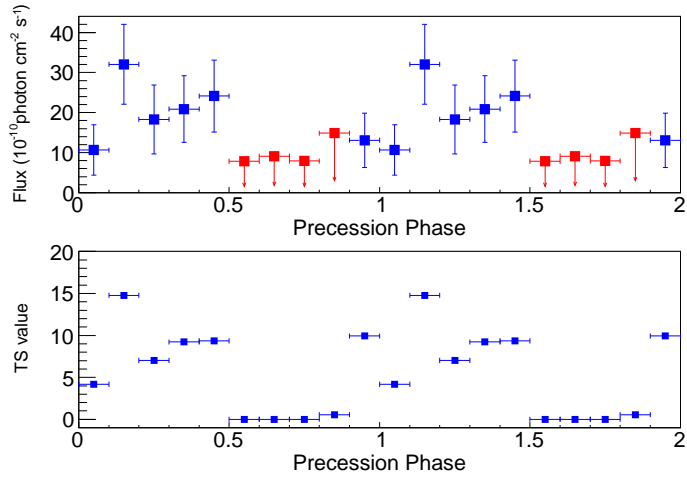}
\caption{Precessional phase light curve (top) and TS values (bottom) of \jj\/ in 1-300 GeV {with a binning of 0.1}.
The upper limits are at the 95\% confidence level.
}
\label{10bin}
\end{figure*}
\end{center}

To calculate the significance of the flux deviation from a constant between precessional phase 0.0--0.5 and 0.5--1.0, the corresponding two data sets were fitted simultaneously using summed likelihood analysis.
The spectral index of \jj\/ was fixed to the value derived from precessional phase-averaged data.
The analysis was carried out first with the normalization of \jj\/ tied together through the two data set, and then repeated with it untied.
The $\Delta$TS between two maximum log-likelihood values is 12, which corresponds to a significance of 3.5$\sigma$ and is consistent with the timing signal.
The binning in the precessional phase, i.e.,  0--0.5 and 0.5--1, is arbitrary and adopted a priori based on the ephemeris.
Thus, only one trial is introduced in the analysis.
Because of the lower exposure time in smaller precessional phase bins, the corresponding uncertainties in the individual flux measurements grow, see Figure \ref{precession}.
However, as the latter Figure shows already, we tested a posteriori that the coarse variability trend is maintained even at smaller bins (e.g., even dividing the precessional phase in 10 bins, Figure \ref{10bin}).

As a consistency check, we carried out the same timing and precessional phase-related  likelihood analysis in the 0.1-1 GeV band.
No significant periodic signal is detected and no flux variation can be claimed.
The $\Delta$TS between maximum log-likelihood values of precessional phase 0.0--0.5 and 0.5--1.0 is 1.6.

The same check was further carried out above 1 GeV for the full dataset, i.e., without pulsar gating of \psrj.
A weak hint of a periodic signal at {161.93$\pm$2.91} days is detected from \jj\, with a single-frequency significance of {2.7$\sigma$}.
The $\Delta$TS between maximum log-likelihood values of precessional phase 0.0--0.5 and 0.5--1.0 is 8, indicating a flux deviation from a constant at 2.8$\sigma$ level.
The decreased significance of timing and flux variation are most likely due to the larger number of events that originate in \psrj\/; its Poissonian flux fluctuation will smear the periodic flux variation from \jj\/, which is much dimmer in comparison.

%%%%%%%%%%%%%%%%%%%%%%%%%%%%%%%%%%%%%%%%%%%%%
\textbf{Stability of the timing signal}
%%%%%%%%%%%%%%%%%%%%%%%%%%%%%%%%%%%%%%%%%%%%%

\begin{center}
\begin{figure*}[!h]
\begin{minipage}[tb]{0.9\textwidth}
\flushright
\includegraphics[scale=0.7]{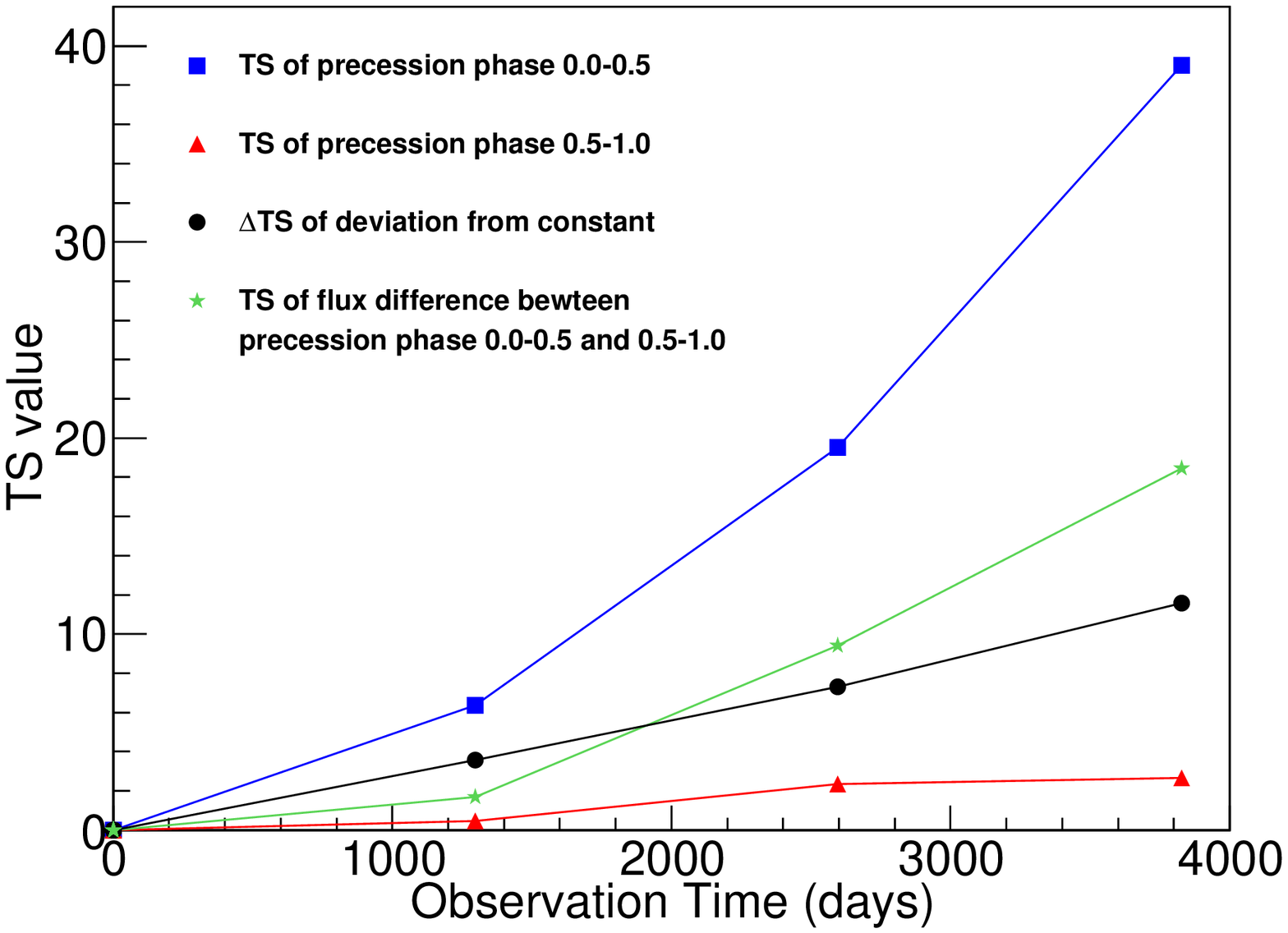}
\end{minipage}

\begin{minipage}[tb]{0.903\textwidth}
\flushright
\includegraphics[scale=0.5, width=14.45cm,height=4.6cm]{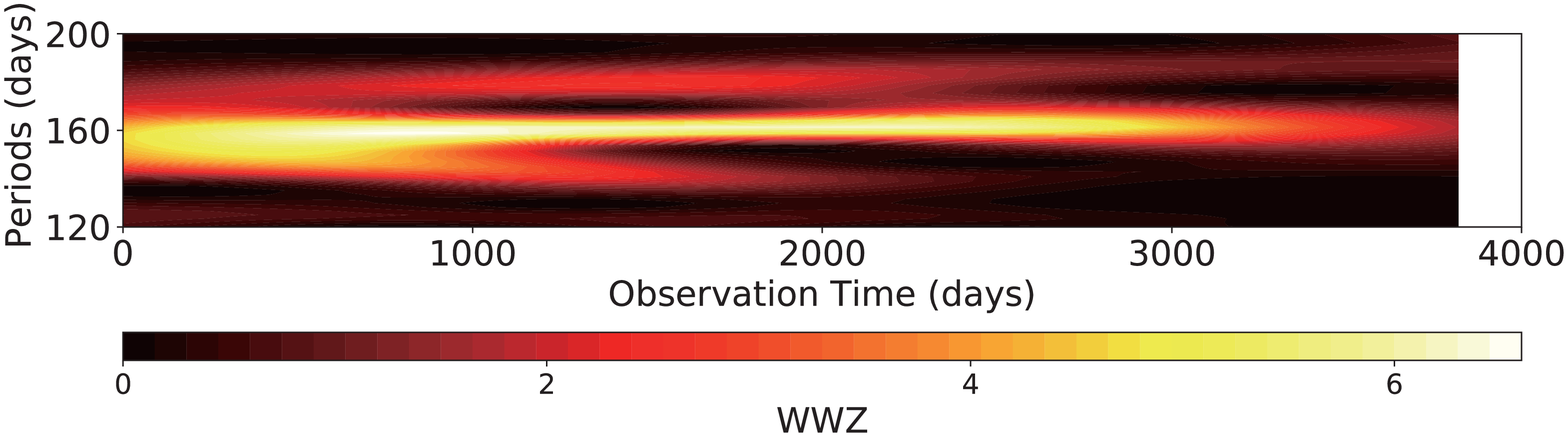}
\end{minipage}
\caption{Top: cumulative likelihood analysis during precession phase 0.0--0.5 and 0.5--1.0.
 the TS of \jj\/, $\Delta$TS of the flux deviation from a constant and TS of the flux difference between two precessional phase bins are shown with different color and marker.
Bottom: 2D plane contour plotting for the WWZ power spectrum.
}
\label{wwz}
\end{figure*}
\end{center}

%
We also carried out a cumulative likelihood analysis during the precession phase 0.0--0.5 and 0.5--1.0 in 1-300 GeV.
To allow for significant measurements along the evolution, we adopted a step of 8 precessional periods (1298 days), and show the
evolution in time in Figure \ref{wwz}, top panel.
The TS of \jj\/ during the precessional phase 0.0--0.5 increases as observation time accumulates, while it stays almost unchanged during the precessional phase 0.5--1.0.
As a result, the flux difference between the two precessional phase bins becomes more significant, providing additional credibility to the timing signal reported.
Additionally, to explore further the stability of the timing signal, we employed the Weighted Wavelet Z-transform (WWZ\cite{Foster1996}).
The 2D plane contour plotting for the WWZ power spectrum is shown in Figure \ref{wwz}, bottom panel.
The timing signal at $\sim$160 days is present along the 10.5 years observation time, but some intensity variation is apparent.
Such variations may be an expected outcome in the scenario described if the injection is not constant, or the magnetic tube is not fixed in space.

%%%%%%%%%%%%%%%%%%%%%%%%%%%%%%%%%%%%%%%%%%%%%
\textbf{{Neutral atomic gas analysis}}
%%%%%%%%%%%%%%%%%%%%%%%%%%%%%%%%%%%%%%%%%%%%%

To compare the gamma-ray emision of SS 433 with the large-scale gas in the region, we used the 21 cm
emission line of H\,{\sc i} as a tracer of the neutral atomic gas.
The Galactic ALFA H\,{\sc i} (GALFA\cite{peek2011,peek2018}) survey data from the Arecibo Observatory 305 m-telescope was investigated.
These data were first used in the report by\cite{Su2018}.
The GALFA H\,{\sc i} cube data have a grid spacing of 1 arcmin and a velocity channel separation of 0.184 km s$^{-1}$.
Typical noise levels are 0.1 K rms of brightness temperature in an integrated velocity of 1 km s$^{-1}$.

Based on the Arecibo 21 cm H\,{\sc i} data, we found an atomic {gas excess} coincident with \jj\/ at V$_{LSR}$~66km/s, which corresponds to a distance of $\sim$ 4.1$\pm$0.7 kpc, consistent with that of SS 433\cite{Gaia2016, Reid2014}.
The {enhancement} is located at R.A. = 288.11$\degree$, decl.= 5.31$\degree$ with a radius of $\sim15$ arcmin ($\sim$20 pc at 4.6 kpc).
The H\,{\sc i} intensity of the main structure is $\sim$ 1000 K km s$^{-1}$, leading to the total mass of $\sim 25000$M$_{\odot}$.
The volume-averaged density of the H\,{\sc i} gas enhancement is estimated to be $\sim$22 cm$^{-3}$, assuming a spherical region of radius $R_c$.
Given the relatively large grid spacing of the observations, the H\,{\sc i} gas enhancement could be located in clumps or in a central cusp.
The existence of clumps has often been found when clouds are observed at higher resolution in our Galaxy\cite{Heyer2015} and beyond\cite{Fukui2010}.
{However, the low average density, on the other hand, make strong clumpiness, albeit in principle possible, unlikely\cite{Bergin2007}.}

\textbf{{X-ray data analysis}}

We have also analyzed two \emph{ROSAT}/PSPC observations of the SS 433 region.
\emph{ROSAT}/PSPC ObsID RP400271A01 provides 20 ks exposure on the east lobe, whereas ObsID RP500058A02 provides 21 ks exposure on the west lobe of SS 433.
The data were analyzed using xselect and ftools as suggested by the \emph{ROSAT} data analysis manual\cite{rosat}.
Exposure corrected X-ray images in 0.9--2.0 keV were produced and found to be consistent with\cite{brinkmann1996}.
The contour of east and west lobes in X-ray were shown in Figure \ref{tsmap}, top panel.

\textbf{{GeV emission due to hadronic interactions?}}

In order to assess whether the GeV emission could be due to hadronic interactions, we considered
a numerical solution of the (isotropic) diffusion equation --from an injection point up to a given distance-- to compute the cosmic ray density.
We then used the latter to compute the gamma-ray emission (similarly to what was done in \cite{Aharonian1996,Bosch2005}).

Provided that the interstellar cosmic-ray proton energy density in the region of SS~433 is the same as the locally detected one,
we find that either in a continuous conversion of a fraction of the jet kinetic luminosity to cosmic rays, or in an impulsive cosmic-ray injection event of much
shorter duration than the age of the jets, assumed here as $\sim 2\times 10^4$ years, as in e.g., \cite{Panferov2017} (albeit the argument would not be significantly changed in case this age is smaller, see \cite{Goodall2011}),
the cosmic-ray density at the cloud could exceed the Galactic sea, generating
gamma rays against the averaged proton density at a comparable level of flux to the one detected.
The average level of gamma-ray flux we measure from \jj\ is equivalent to a luminosity of $10^{34}\rm erg~s^{-1}$ at the SS 433 distance.
To accommodate this luminosity via hadronic interactions, we need a total  $W_p=2.5\times 10^{48}(L_\gamma/10^{34}\rm erg~s^{-1})(n/20\rm cm^{-3})^{-1}$ erg in cosmic ray protons interacting with the atomic cloud.
Protons might be accelerated at the terminal lobe, or come from the SS 433 equatorial outflow, and reach \jj\ through isotropic diffusion.
The distance from east termination lobe or from the SS 433 central object to \jj\ are similar (See Figure \ref{tsmap}).
The kinetic power of both the jet and the equatorial outflow are also similar $\sim 10^{39}\rm erg~s^{-1}$\cite{Middleton2018}.
Thus, in either scenario an accumulation of proton injection over $\sim 100\,$yr is needed to supply the required proton energy, and consequently, any periodical signal due to injection will be smeared out.

Even if assuming that a sufficient cosmic-ray energy is injected in each single period to account for the observed gamma-ray flux, the difference of the arrival time to the cloud of cosmic rays between two consecutive injection events separated by a precessional period is small in comparison to the precessional period itself, such that many injections would still accumulate in the cloud,  erasing the period signal produced by the arrival of fresh protons.

\begin{center}
\begin{figure*}
\centering
\includegraphics[scale=0.3,angle=-90]{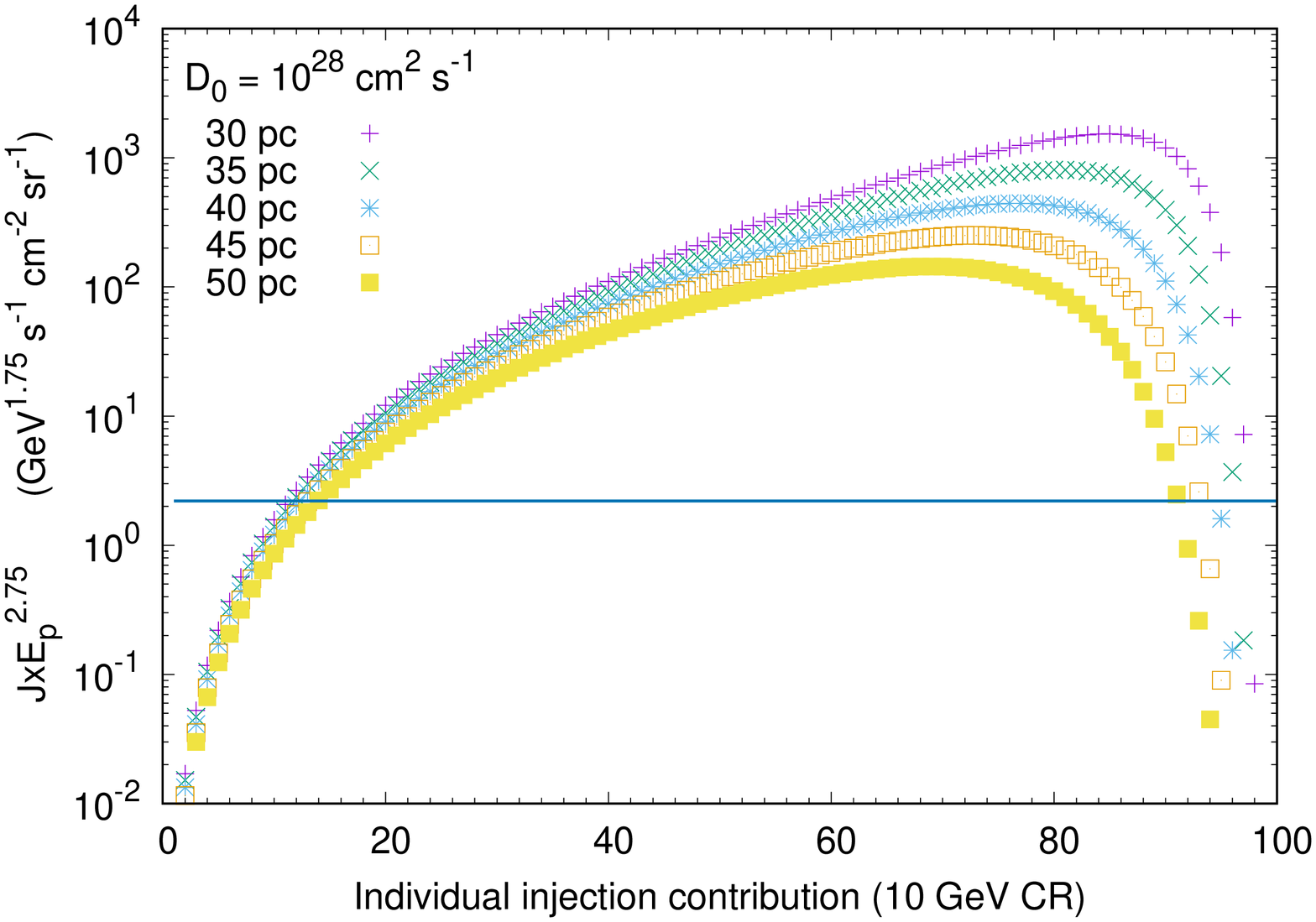}
\includegraphics[scale=0.3,angle=-90]{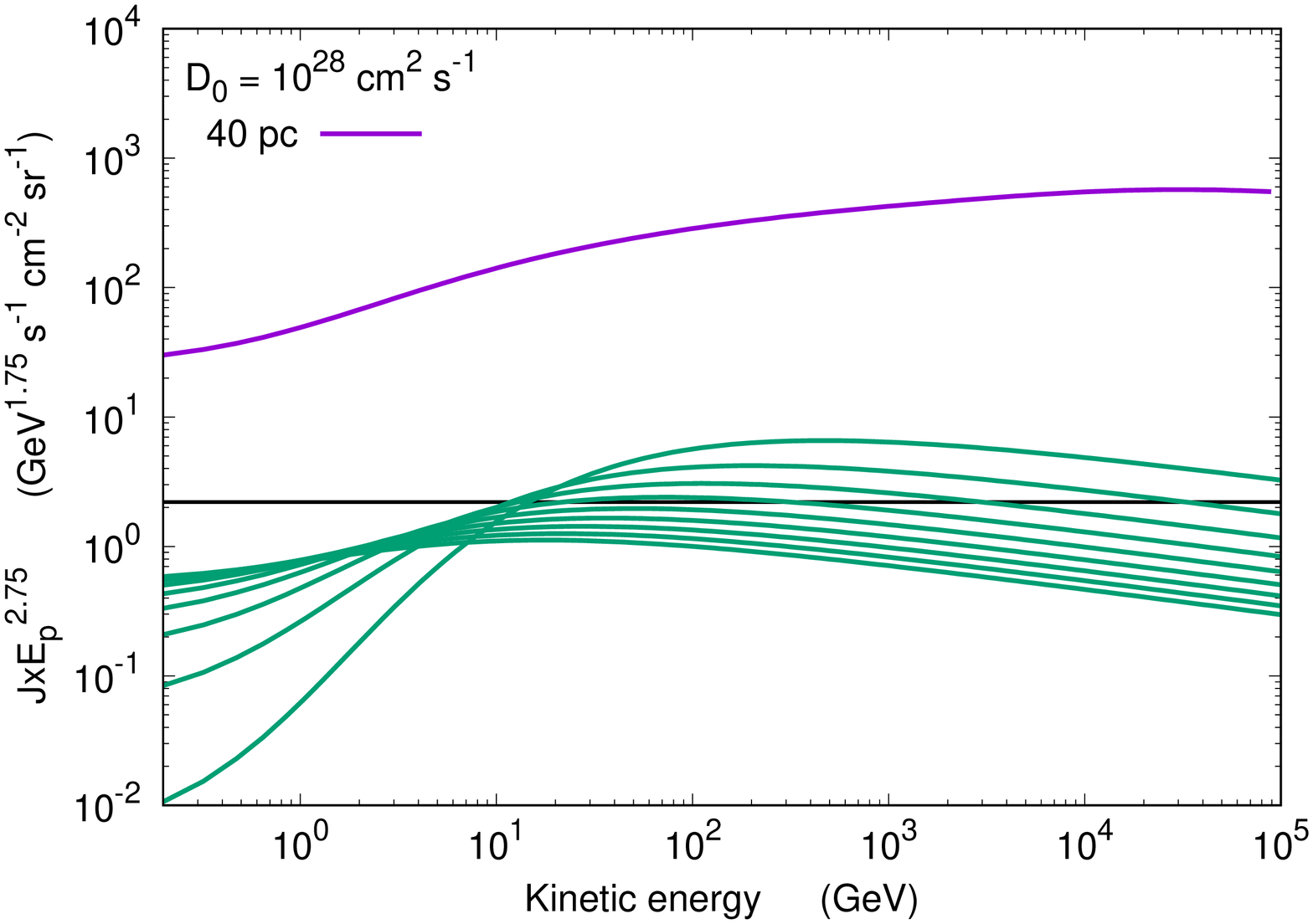}
\includegraphics[scale=0.3,angle=-90]{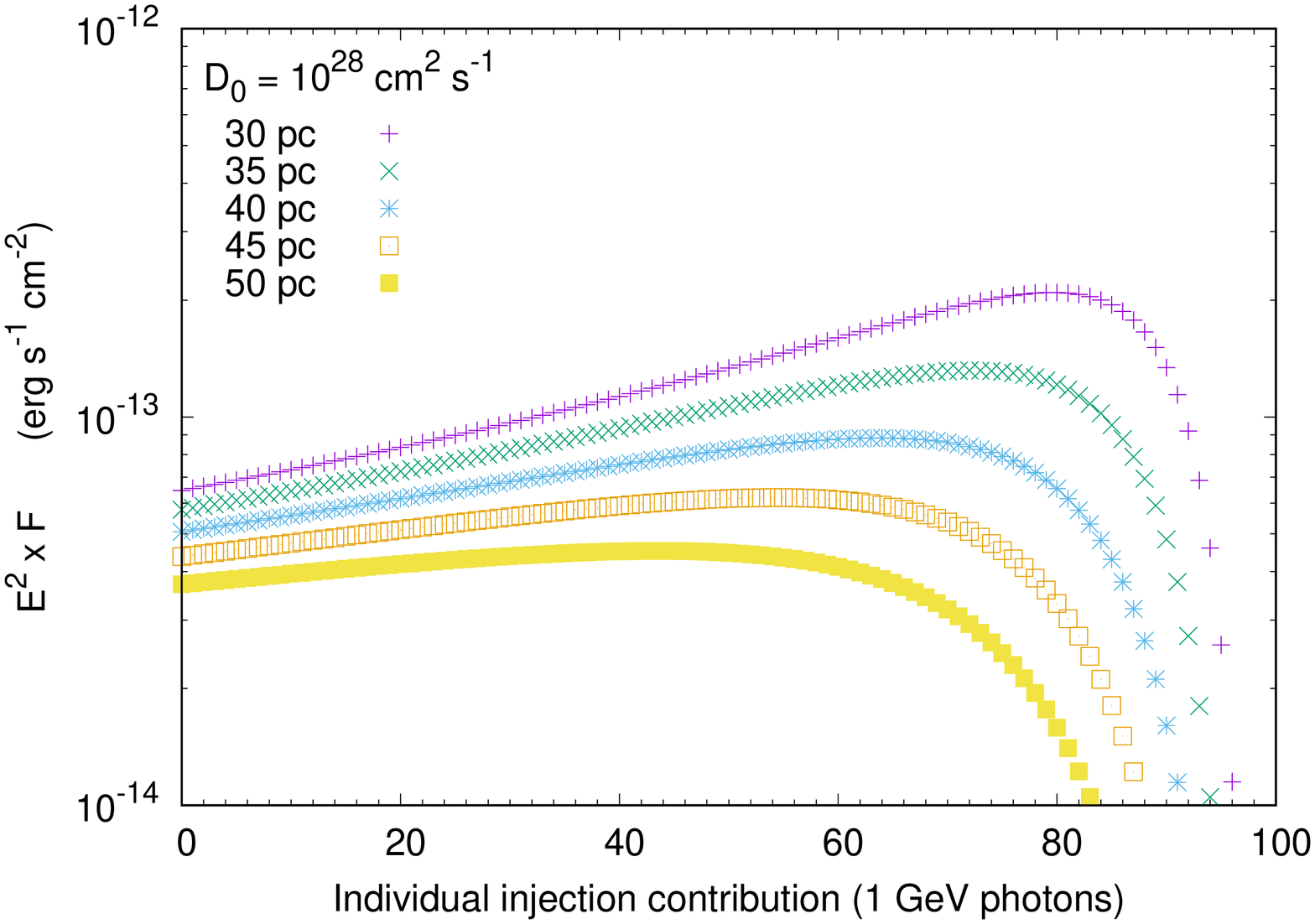}

\caption{Examples of simulations for a periodic, instantaneous injection of protons. Details are described in the text.
}
\label{ss433-diff}
\end{figure*}
\end{center}

\begin{center}
\begin{figure*}
\centering
\includegraphics[scale=0.6,angle=-0]{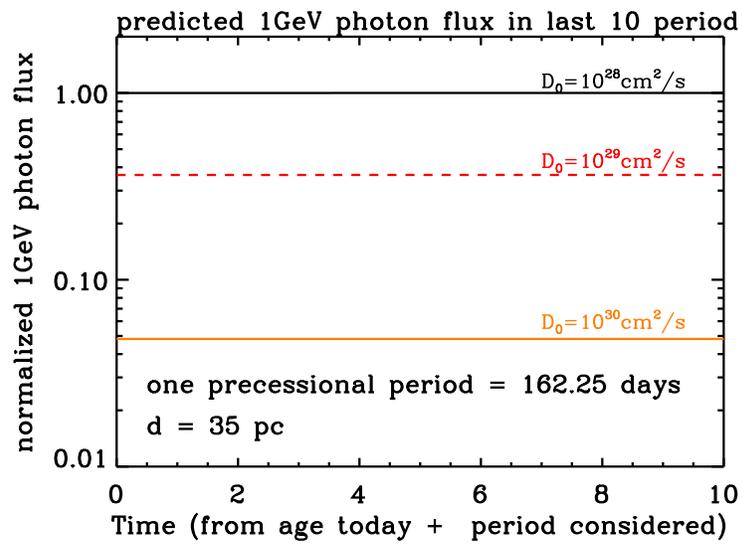}

\caption{Predicted 1 GeV gamma-ray flux in the last ten precessional periods computed with different diffusion coefficient. Details are described in the text.
}
\label{ss433-diff-2}
\end{figure*}
\end{center}

This situation is quantitatively exemplified in Figure \ref{ss433-diff} where we are purposely considering a mock period 500 times larger than the precessional one of \sss\ so that both, each individual instantaneous injection contains a larger amount of energy (equal to the kinetic power released in the period) and is more numerically tractable (given the need of considering only 100 periods to cover a significant age $\sim 22000$ yrs).
In order to maximize even further the cosmic-ray luminosity at the cloud, this example will
also purposely consider a high total luminosity of $3\times10^{40}$ erg s$^{-1}$, of which 20\% is assumed to end in cosmic rays at the injection point.
We assume that the latter propagate isotropically in a medium with diffusion coefficient of 10$^{28}$ cm$^2$ s$^{-1}$; we tested that
changes in the latter will not modify conclusions.
The first panel of Figure \ref{ss433-diff} shows the contributions to 10 GeV cosmic-ray (an example of an energy relevant for producing 1 GeV photons) at different distances from the injection point (around the separation between the cloud and \sss) of 100 individual injection events, compared to the Galactic cosmic-ray sea -represented by the horizontal line.
When the injection is very old, protons diffuse into a volume much larger than the region of interest.
On the contrary, when it is too fresh (the last injection events, towards 100 in the x-axis of the figure), cosmic rays have not yet arrived to the region of interest.
In both cases, the contribution to the total cosmic-ray density at that energy is low.
Instead, in intermediate times, each injection event can provide more 10 GeV cosmic rays than the Galactic sea.
Note that this would not be the case should the real precessional period of \sss\ be considered (500 times smaller), i.e., the individual injections would be sub-dominant to the Galactic sea in that case although the general scenario would be maintained vis-a-vis.
The second panel of Figure \ref{ss433-diff}  shows the cosmic-ray density (at all energies): the green lines represent 10 individual
injections separated (corresponding to 10, 20, 30... 100 in the x-axis of the first panel), whereas the violet line shows the sum of the contribution of all injection events.
Similarly, the third panel of Figure \ref{ss433-diff}  shows the hadronic gamma-ray emission at 1 GeV (obtained from a computation of a full spectral energy distribution, with the corresponding proton density) at different distances.
To exemplify this further, we show in Figure \ref{ss433-diff-2} how the GeV gamma-ray flux evolves in time within the last ten processional periods after an injection of $\sim 22000\,$yrs. Again, we consider an impulsive periodical injection of cosmic rays and isotropic diffusion, but here we use with the real jet precessional period 162.25\,days. The cloud is located 35\,pc away from the cosmic-ray injection point.
For typical ISM diffusion coefficients (i.e., $D_0 \sim 10^{28}\rm cm^2s^{-1}$), we cannot see any hint of the periodicity in gamma-ray flux, with a complete loss of the injection memory. We also examine the influence of the diffusion coefficient. A larger diffusion coefficient would in principle help to reveal the periodicity. However, we still cannot find any periodicity even with a diffusion coefficient up to $D_0=10^{30}\rm cm^2s^{-1}$.
To focus on the temporal behavior, we normalize the fluxes in all three employed diffusion coefficients to the flux in the case of $D=10^{28}\rm cm^2 s^{-1}$. We can see the resulting gamma-ray flux decreases with increasing diffusion coefficient. So even if a larger diffusion coefficient, with which the periodicity can be present, is somehow achieved, it would lead to an energy budget crisis to power the observed gamma-ray flux.
Thus, this approach would be unable to explain the periodicity.

There are two ways of conceptually alleviating this situation.
One can consider that cosmic rays interact with a gas enhancement in a small clump(s) or cusps of density.
This helps in maintaining coherence of the signal when cosmic rays arrive periodically.
One can also consider that
anisotropic diffusion or streaming of cosmic rays along a flux tube is active, so as to ease the energetics:
Differently to isotropic diffusion, streaming along a slim tube connecting the gas cloud and the particle injection point could in principle allow for most cosmic rays produced to arrive
at the region of interest.
In this way, if most protons injected within one precessional period arrive at the cloud clump, and quickly
exit in half a period (i.e., $\sim$80~days or $7\times 10^6$s, and a size of $R<c\times 80\rm ~days\sim 0.07~pc$ for the cloud clump), periodic coherence can in principle be maintained in gamma-rays.
Using the mean free path for hadronic interactions ${\rm mfp} \sim (n \sigma_{pp})^{-1}$, with $n$ the clump density and $\sigma_{pp}$ the hadronic cross section,
we can see that assuming a clump density of about $\sim 10^4 \rm cm^{-3}$,
the probability for them to interact is $\sim (1-\exp(-(R/2)/{\rm mfp} )) \sim 10^{-4}$.
If $R$, the clump size is about 0.07 pc (so that $R/2$ is about the light crossing time), with the assumed clump density the total mass in the cloud is quite small, $\sim 1$M$_\odot$;
acceptable from the observational perspective (overall values of average density, mass, size, and angular resolution of observations).
With this mass, considering that 10\% of the power in the outflow (taken as $3\times 10^{39}$ erg s$^{-1}$) goes into cosmic rays in each periodic illumination (i.e., assuming that 10\% of the power in the outflow along half a precession period goes into cosmic rays), and the probability for them to interact, the level of flux resulting from each injection event is comparable to observations.
Also, note that each injection the magnetic flux can impact different regions of the cloud, what would explain some variability in the periodically produced signal.
Despite how unlikely it might be to have a magnetic streaming connecting the injection point with the cloud, the low mass needed for the large energetics does not allow to rule this idea out, leaving it open for future studies.

\textbf{{Data Availability}}

The datasets generated during and/or analysed during the current study are available from the corresponding author on reasonable request.
\newpage

\textbf{References}
\bibliography{ss433}

\label{results}
%%%%%%%%%%%%%%%%%%%%%%%%%%%%%%%%%%%
%%%%%%%%%%%%%%%%%%%%%%%%%%%%%%%%%%

%%%%%%%%%%%%%%%%%%%%%%%%%%%%%%
%%%%%%%%%%%%%%%%%%%%%%%%%%%%%%%%%%

\vspace{-3.5em}
%%%%%%%%%%%%%%%%%%%%%%%%%%%%%%
%%%%%%%%%%%%%%%%%%%%%%%%%%%%%%%%%%

%%%%%%%%%%%%%%%%%%%%%%%%%%%%%%
%%%%%%%%%%%%%%%%%%%%%%%%%%%%%%%%%%

\vspace{-3.5em}
%%%%%%%%%%%%%%%%%%%%%%%%%%%%%%
%%%%%%%%%%%%%%%%%%%%%%%%%%%%%%%%%%

\vspace{-3.5em}
%%%%%%%%%%%%%%%%%%%%%%%%%%%%%%
%%%%%%%%%%%%%%%%%%%%%%%%%%%%%%%%%%%

\end{document}